\documentstyle[prl,aps,epsf]{revtex}
\begin{document}
\twocolumn[\hsize\textwidth\columnwidth\hsize\csname %
@twocolumnfalse\endcsname

\draft
\title{Signatures of Stripe Phases in Hole 
Doped La$_{2}$NiO$_{4}$}
\author{Ya-Sha Yi, Zhi-Gang Yu, A. R. Bishop and J. Tinka Gammel}
\address{
Theoretical Division and Center for Nonlinear Studies, 
Los Alamos National Laboratory, Los Alamos, NM 87545}
\date{November 21, 1997}
\maketitle
\begin{abstract}
We model nickelate-centered and oxygen-centered 
stripe phases in doped La$_{2}$NiO$_{4}$ materials. We use an  
inhomogeneous Hartree-Fock and random-phase approximation 
approach including both electron-electron and electron-lattice(e-l) 
coupling for a layer of La$_{2}$NiO$_{4}$. We find that whether 
the ground state after commensurate hole doping comprises  
Ni-centered or O-centered charge-localized stripes  
depends sensitively on the e-l  
interaction. With increasing e-l interaction 
strength, a continuous transition from an O-centered stripe phase 
to a Ni-centered one is found. Various low- and high-energy 
signatures of these two kinds of stripe phases are predicted,  
which can clearly distinguish them. These signatures 
reflect the strongly correlated spin-charge-lattice features  
in the vicinity of Ni-centered or O-centered stripe domains.  
The importance of e-l interaction for recent experiments 
on stripe phases is discussed. 
\end{abstract}
\phantom{.}

]

\narrowtext
\pagebreak

\section{Introduction}
The interplay of spin, charge and lattice degrees of freedom 
has been the focus of  
many classes of electronic materials recently, and results 
in numerous novel properties\cite{1,2,3,4,5}.  
An important example lies in the growing appreciation for 
the presence and importance of charge localization(small polaron 
formation) and ordering 
tendencies in transition metal oxides. Stripe phases are suggested 
\cite{6} in a variety of nickelates, cuprates and manganites, 
i.e, fine-scale domain walls separating spin, charge and lattice 
degrees of freedom into superlattice and long-period patterns. 
The likely importance of charge coupling to the 
lattice is becoming evident in determining the 
formation of stripes when doping into a background of 
antiferromagnetic order\cite{7,8,8b}: 
electron-lattice(e-l) interaction strongly assists localization 
of the holes through a positive feedback mechanism.

The possible existence of such 
hole-rich stripe phases has been proposed in the form of 
antiphase domain 
boundaries between antiferromagnetically ordered spins in the 
study of some high temperature superconducting oxides
\cite{9,10,11,12,13,14} and colossal magnetoresistance manganites\cite{14a}. 
Stripes were first observed in nickelate oxides\cite{15}, and 
the Hubbard model in Hartree-Fock(HF) approximation 
gave a good description of the 
ground state of the doped nickelates\cite{7}.  
Charge ordering is the driving force for the 
formation of stripes\cite{16}. One feature of both theoretical and 
experimental interest concerns the orientation of the stripes 
relative to the lattice\cite{8,17}. Specifically, whether the stripe in 
nickelate oxides is Ni-centered or O-centered, and the possible 
mixture of these two kinds of stripes, has been investigated 
intensively in recent experiments\cite{8,17}. 
Through magnetic-field-induced 
effects, it is suggested that 
the stripe phase is O-centered when the temperature T is 
higher than a temperature T$_{m}$, while the 
Ni-centered stripe phase is favored when T is lower than T$_{m}$. However,   
it is difficult to unambiguously determine the stripe form 
directly in a standard neutron or X-ray scattering 
experiment because of the loss of phase information.  
Similar ideas of  
site-centered or bond-centered stripe phases in cuprate oxides are  
also under investigation\cite{17b}.  

Previous modeling of the nickelates gave a reasonable explanation 
of the mechanism for the formation of stripe phases, but no detailed 
studies on the experimentally observable consequences has been given. The 
Ni-centered and O-centered stripe phases  should certainly 
exhibit different experimental signatures.  Due to the 
difficulty of  current scattering experiments to directly 
determine the stripe form, it is very important to investigate 
these distinguishing observable signatures. As in other small-polaron type 
phenomena, the relationship between high energies(e.g. optical 
absorption) and low energy(e.g. lattice spectroscopy) can serve 
to specify particular structures very distinctly\cite{22}. 

For a comprehensive understanding of the competition among 
the spin, charge and lattice degrees of freedom in 
complex electronic materials such as 
nickelate oxides, specifically for the spatially 
inhomogeneous stripe patterns within models of NiO$_{2}$ planes, we 
resort here to the inhomogeneous HF plus RPA approach\cite{18,19}. 
This has proven to be a robust method for such 
phenomena, especially when electron-lattice coupling 
is present, obviating subtle many-body effects and quantum 
fluctuations\cite{7,20}. Using a 2-dimensional(2D) 
tight-binding Peierls-Hubbard model Hamiltonian, we obtain both  
Ni-centered and 
O-centered stripe phases.  
The Ni-centered stripe phase is energetically  favored at strong 
e-l interaction, while the O-centered stripe phase is  
stable in the weak e-l coupling regime. The transition from the 
O-centered stripe phase to the Ni-centered one can be  
clearly identified by defining an appropriate "order 
parameter", as will be discussed 
in detail below. {\it Local} phonon, charge and magnetic 
 modes are found 
for both of the two kinds of stripe phases, strongly distinguishing 
the inhomogeneous ground states from the homogeneous 
undoped situation. The infrared(IR) and Raman spectra and the 
magnetic excitations for both the nickelate spin and the oxygen spin 
are distinctive signatures, which should serve to guide 
experiments to discriminate between Ni-centered  or O-centered stripes.  
Likewise, there are HF levels induced by stripes in the electronic 
charge-transfer-gap of the undoped case(i.e.intra-gap levels), 
which are reflected in the optical absorption. 

This paper 
is organized as follows. In Sec.II our 2-dimensional(2D) Hamiltonian 
model for an NiO$_{2}$ plane is 
given,  and its various parameters are presented. The differences 
between this model Hamiltonian for the nickelates and the corresponding 
one for cuprates\cite{9} 
are emphasized. The Ni-centered and the O-centered stripe phases 
and their respective characteristics are comparered in 
Sec.III. In Sec.IV we discuss predicted observable signatures 
for both Ni-centered and O-centered stripe phases. A summary and 
remarks are presented in Section V.  

%\newpage

\section{Model Hamiltonian and Parameters}
We consider a 2D four-band extended Peierls-Hubbard model, which 
includes both  electron-electron and electron-lattice interactions 
\cite{18,19}. Here, for nickelate oxides, besides the 
$d_{x^{2}-y^{2}}$ orbital usually included in cuprate 
oxide models\cite{9,18,19},  the $d_{3z^{2}-1}$  
Ni $d$ orbitals must be included to account for the high spin 
state($S=1$) at half filling(i.e. undoped). The model 
Hamiltonian we study is: 

\vskip 6pt

\begin{equation}
\begin{array}{lllll}
H=&\displaystyle\sum\limits_{i\neq j,m,n,\sigma}t_{ij,mn}({u_{k}})c_{im\sigma}^{\dagger}
c_{jn\sigma}+\sum\limits_{i,m,n,\sigma}e_{imn}({u_{k}})c_{im\sigma}^{\dagger}
c_{im\sigma} \\
  &\displaystyle +(U+2J)\sum\limits_{i,m}n_{im\uparrow}n_{im\downarrow}+
(U-\frac{1}{2}J)\sum\limits_{i,m\neq n,\sigma,\sigma^{'}}n_{im\sigma}
n_{in\sigma^{'}} \\
&\displaystyle -2J\sum\limits_{i,m\neq n}{\bf S_{i,m}}\cdot {\bf S_{i,n}} 
+J\sum\limits_{i,m,n}c_{im\uparrow}^{\dagger}c_{im\downarrow}^{\dagger}
c_{in\downarrow}c_{in\uparrow} \\
&\displaystyle +\sum\limits_{i\neq j,m,\sigma,\sigma^{'}}U_{ij}n_{im\sigma}n_{jn\sigma^{'}}
+\sum\limits_{l}\frac{1}{2M_{l}}p_{l}^{2} \\
&\displaystyle+\sum\limits_{k,l}\frac{1}{2}K_{kl}u_{k}u_{l} . \\
\end{array}
\end{equation}
Here, $c_{im\sigma}^{\dagger}$ creates a hole with spin $\sigma$ at  
site $i$ in the Ni $d_{x^{2}-y^{2}}$, $d_{3z^{2}-1}$ or O $p_{x,y}$ 
orbital. The electron-electron interactions include the Ni-site($U_d$) 
repulsions  and the O-site($U_{p}$) repulsions for $U_{i}$, as well as 
the Hund interaction $J$ at the same Ni site to account for  
the high spin situation.(The interplay of the two orbitals can 
also lead to effects such as pseudo Jahn-Teller distortions. 
These are not our main focus here but will be mentioned 
briefly below). The nearest-neighbour Ni-O interaction 
$U_{ij}$ is also included. For the lattice terms, we study only 
the motion of O ions along the Ni-O bonds---other oxygen(or Ni) 
distortion modes can be readily included in this approach if necessary. 
For the electron-lattice  interaction, 
we consider that the nearest-neighbour Ni-O hopping is 
modified linearly by the O-ion 
displacement $u_{k}$ as $t_{ij,mn}({u_{k}})=t_{pd}\pm\alpha u_{k}$, where 
the +(-) applies if the bond shrinks(or stretches) with positive $u_{k}$.  
The Ni-site energy is assumed to be modified by the O-ion displacements as 
$e_{imn}({u_{k}})=\epsilon_{d}+\beta\sum_{k}(\pm u_{k})$, where the sum runs 
over the four surrounding O ions.(Other electron-lattice couplings, 
including oxygen buckling, can be readily included in the 
present formalism). Other electronic terms are  
O-O hopping($t_{pp}$) for $t_{ij}$, O-site energy ($\epsilon_{p}$) for 
$e_{imn}({u_{k}})$, with $\Delta=\epsilon_{p}-\epsilon_{d}$.  
We  emphasize that, due to the large spin at a nickel  
site, Hund's rule leads to a  ferromagnetic 
exchange coupling $-2J$, and ${\bf S}_{i,m}=\frac{1}{2}
\sum_{\tau,\tau^{'}}c_{im,\tau}^{\dagger}{\bf \sigma}_{\tau,\tau^{'}}
c_{im,\tau^{'}}$, with ${\bf \sigma}$ the Pauli matrix.  

It is  known that for the nickelate oxides the 
e-l coupling is stronger than in cuprates oxides\cite{7,8}, 
and  is therefore 
likely to play an even more decisive 
role in the formation and nature of stripe phases. Complementary to 
recent studies on 
the 2D 3-band Peierls-Hubbard model in various parameter regimes\cite{19}, 
in this paper we concentrate on the variation of the stripe phase with 
respect to the e-l coupling constants in model (1). We adopt the following 
representative parameters for the nickelate 
materials\cite{7}: $t_{pd}=1, \Delta=9, U=4, 
J=1$, and $K=32t_{pd}/\AA^{2}$ (all in unit of $t_{pd}$). 
In real materials\cite{7,23}, $t_{pd}$ is in the range 1.3eV$\sim$1.5eV. 
Detailed analysis of the Ni-centered 
and the O-centered stripe phases we obtained, as well as their 
sensitivity to some of  
the parameters, will be discussed in the next section. 

To emphasize that different kinds of stripes have individual spectroscopic 
signatures, we also consider the linear spin, charge and lattice 
fluctuations 
around the mean-field states: an O-centered stripe state and 
a Ni-centered stripe state\cite{19}. As the electron-hole  
and the phonon excitations are all included, it is very convenient 
to calculate various physical properties through an RPA 
approach; this also serves as a local stability test of the mean-field 
states.

%\newpage

\section{Ni-centered and  O-centered stripes} 
The inhomogeneous HF states are obtained as solutions of 
self-consistent nonlinear equations without 
a priori assumption of the spin, charge and lattice configurations. 
As shown previously 
\cite{7,19},  adding holes in an undoped antiferromagnetically 
ordered system leads to a strong positive feedback: the domain wall 
is the result of accumulation of holes; conversely, more 
localized holes  distort the lattice more and hence 
the system  gains energy through a larger lattice distortion, balanced 
by the cost in electronic energy,  
around the stripe region. As we will show in the following, this 
positive feedback is decisive in forming the anti-phase domain walls which 
divide the whole system into hole-rich striped regions, within which  
the AF order is suppressed,  and 
hole-poor regions, in which  the AF order is maintained.   
Here we study a 5x5 unit cell system  
or a smaller 3x3 system for the later RPA calculation. 
Periodic boundary conditions are used. To emphasize the 
properties of ideal stripes, we concentrate here on 
the commensurate hole doping case of 5(3) holes in the 5x5(3x3) 
unit cell case. Results for other doping situations will be 
given elsewhere. 

\begin{figure}
\vskip -10pt
\centerline{
\epsfxsize=7.0cm \epsfbox{empty.eps}}
\centerline{
\epsfxsize=7.0cm \epsfbox{empty.eps}}
\vskip -10pt
\caption{Real space distribution of spin and charge 
density for an oxygen-centered stripe phase (a) and a nickel-
centered stripe phase (b). The radius of the circle represents 
the amplitude of the charge density and the length of the 
bars represent the amplitude of the spin density. The 
parameters are $t_{pd}=1, \Delta=9, U=4, J=1$, and $K=32t_{pd}/\AA^{2}$. 
$\alpha=2.0$ for (a) and $\alpha=3.6$ for (b).}
\end{figure}

\begin{figure}
\vskip -10pt
\centerline{
\epsfxsize=7.0cm \epsfbox{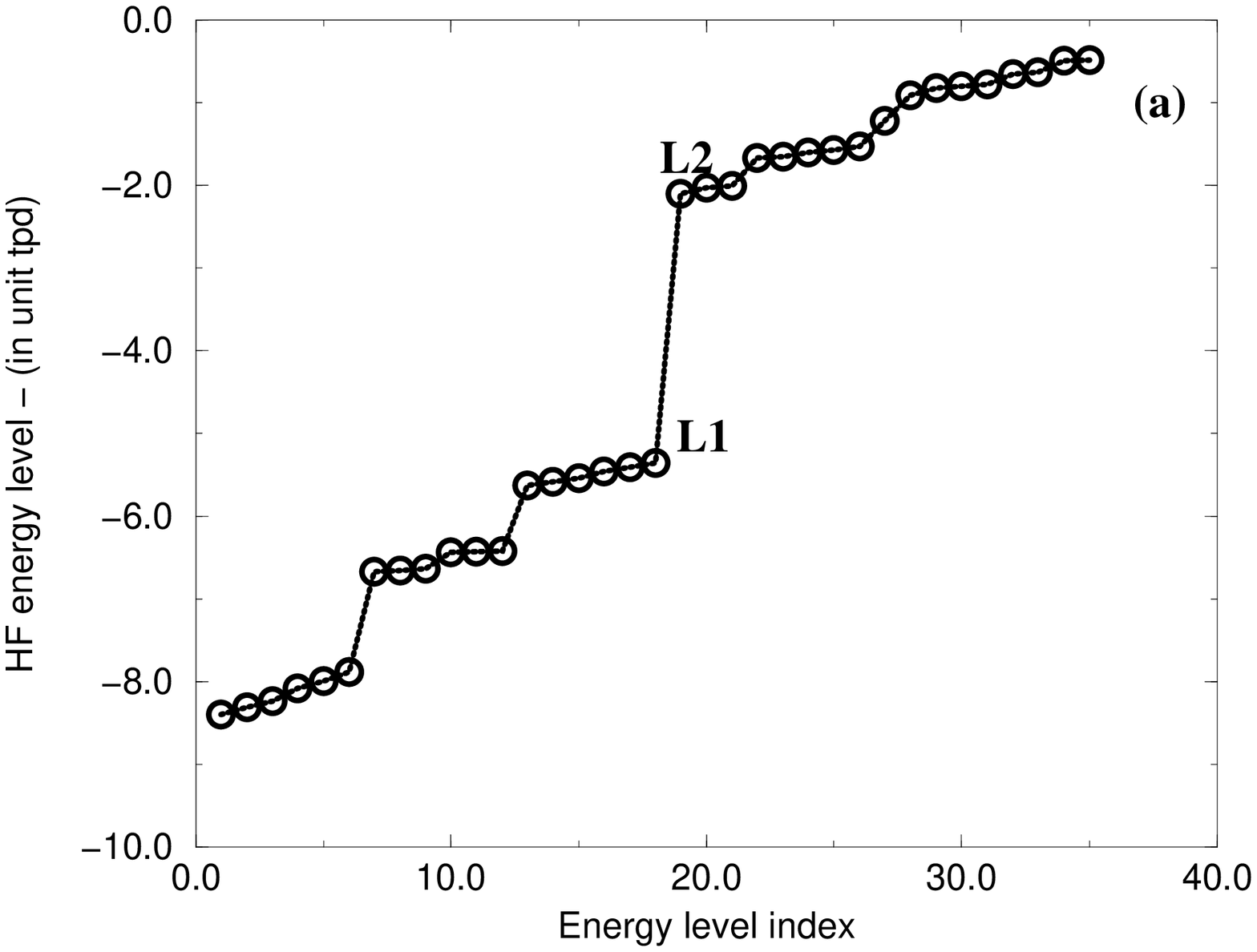}}
\centerline{
\epsfxsize=7.0cm \epsfbox{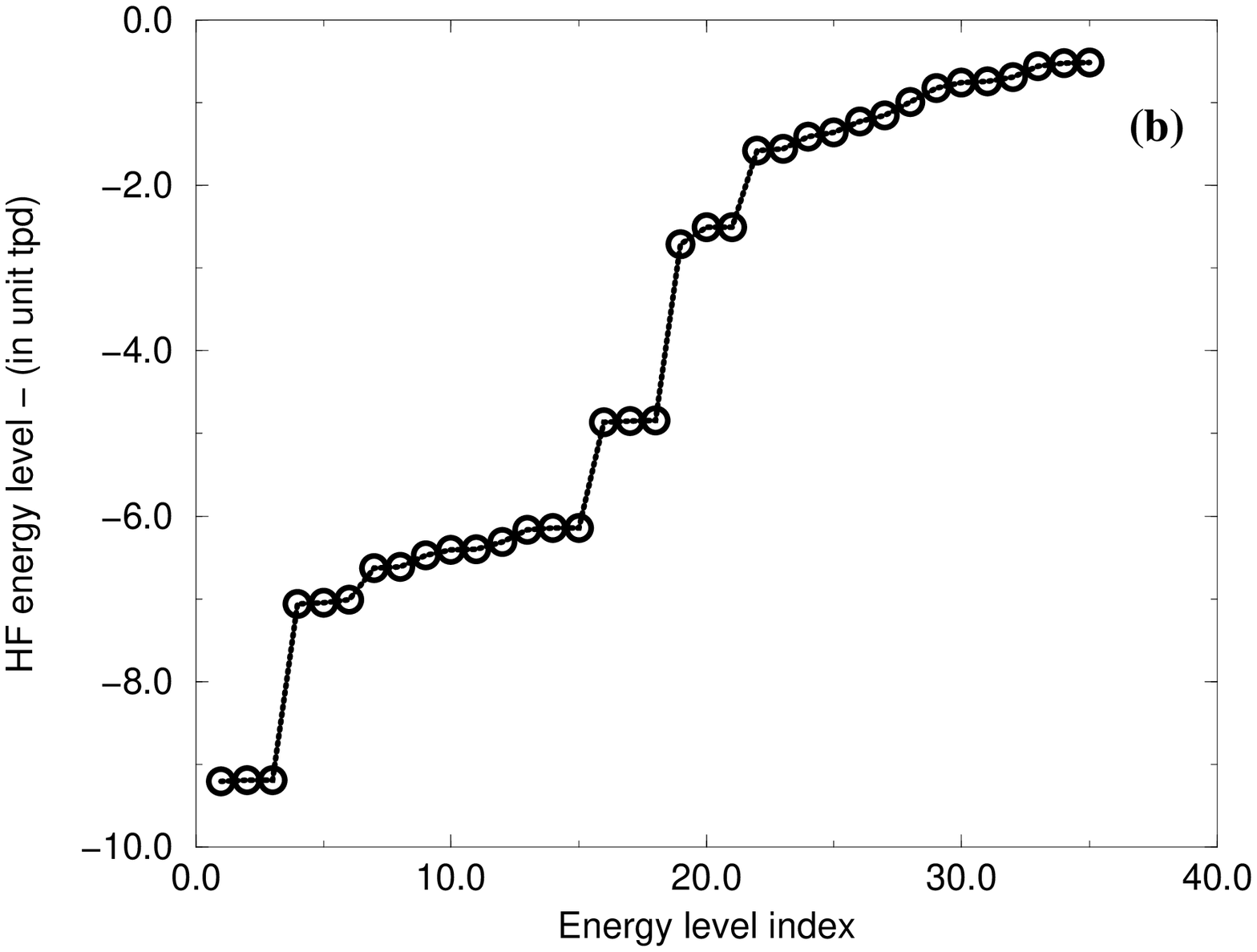}}
\centerline{
\epsfxsize=7.0cm \epsfbox{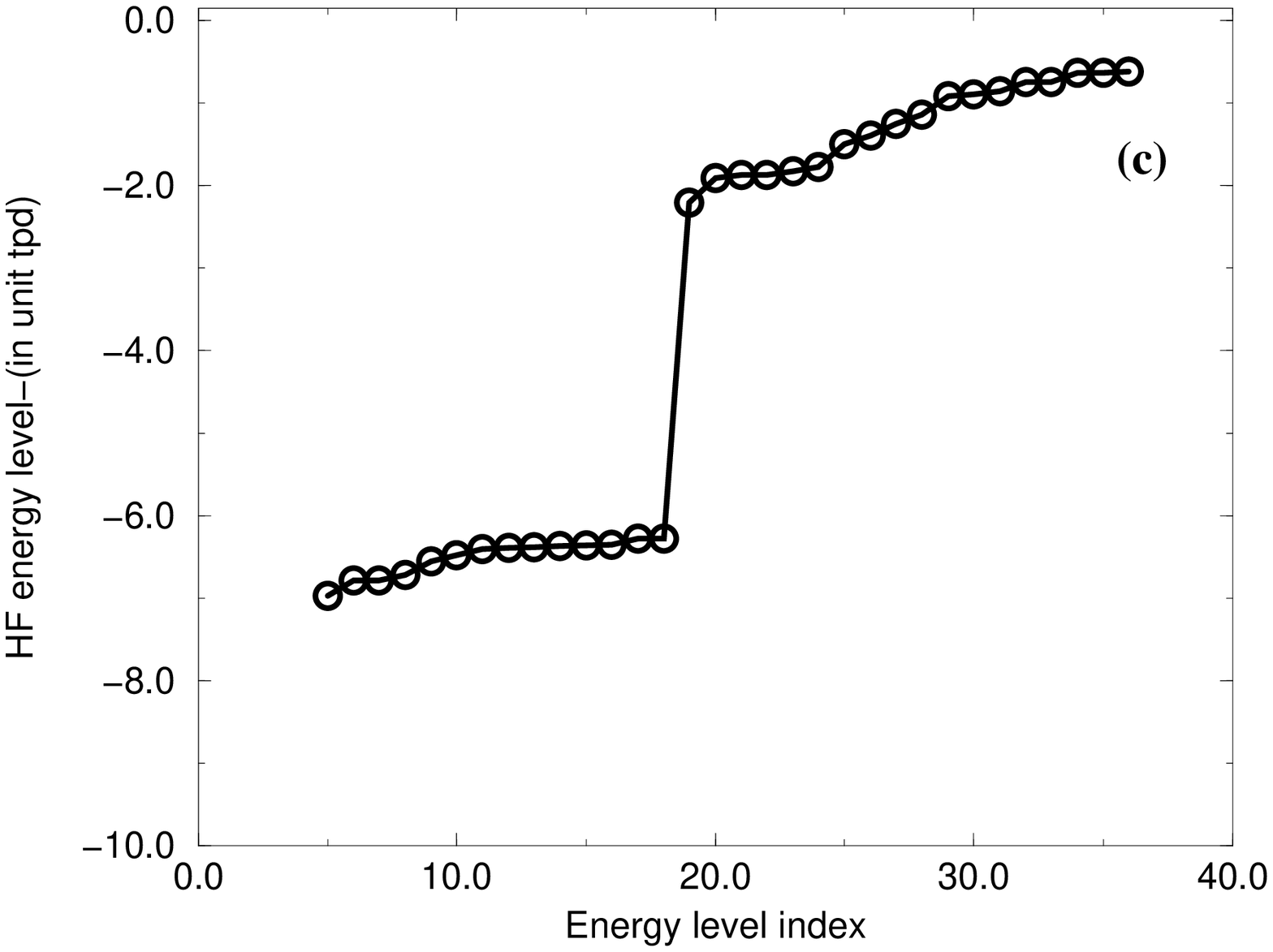}}
\vskip -10pt
\caption{HF energy levels for an O-centered stripe phase (a),  
a Ni-centered stripe phase (b) and undoped case (c).}
\end{figure}

Depending on the e-l coupling strength, 
both  oxygen-centered  and  nickel-centered 
stripe phases are obtained, as illustrated in Figs.1a and 1b, 
respectively. The HF energy levels for these two kinds of 
stripe phases are shown in Figs.2a and 2b(the hole filling 
is to the 21st level), respectively, and the undoped 
case(the hole filling is to the 18th level) is shown in Fig.2c, 
for a smaller 3x3 system, which is for the later comparison 
with the RPA calculation.  
When the holes are doped into the 
system, new localized states appear(as shown 
in Figs.2a and 2b) within the 
charge-transfer-gap of the undoped case(shown in Fig.2c). The 
gap states labelled L1 and L2 for the O-centered stripe phase 
in Fig.2a are most strongly localized. 
A similar conclusion can be also drawn for the 
Ni-centered stripe phase in Fig.2b. From Figs.2a and 2b, we clearly 
see the superlattice gaps induced by the formation of 
stripes, as illustrated in Figs.1a and 1b.    

We found that the O-centered stripe phase is energetically favored 
at weak e-l coupling, while the Ni-centered one is  
favored at strong e-l coupling. This dependence on the electron-lattice 
interaction has been established by using a range of initial 
conditions and confirming local stability through RPA fluctuations. 
We varied the e-l coupling strength  over a 
wide range. The transition from the O-centered stripe 
phase to the Ni-centered one is shown in Fig.3a. In the weak 
coupling region, the O-centered stripe is always found to be  
more stable than 
the Ni-centered one. With e-l coupling constant $\alpha$ increasing beyond 
a critical value($\alpha\simeq 2.2$ with our parameters), 
 the charge 
of holes on oxygen sites along the stripe begins to decrease and 
gradually transfer to the neighboring nickel sites. We also find that 
the  spin 
density on neighboring nickel sites is gradually  
quenched. After the transition region($\alpha\geq 3.0$ for our parameters), 
the charge 
and spin features on the neighboring nickel sites display a fully  
Ni-centered 
stripe phase. The reason for this transition can be understood 
by the competition between the energy gained from the electronic 
kinetic energy and lost from the lattice potential energy,  
with stronger e-l coupling, the system can gain 
much more energy through the electronic kinetic 
energy from the larger lattice distortion; because 
there are 4 oxygens neighbouring  
around each nickel site, the Ni-centered stripes are more 
favored in the strong e-l coupling regime.   

As shown in Fig.3a, there is a small energy barrier 
for the 5x5 system in the  transition 
region between the O-centered stripe phase and the Ni-centered one. 
For our parameters, this energy barrier is about 10K per unit cell. 
However, this barrier is a finite size effect; to elucidate 
this point, we found that in a 7x7 system, the energy barrier is absent, 
while the general trend and the critical point for the transition 
from an O-centered stripe 
to a Ni-centered one is essentially the same.    

The average variations of both spin and charge densities on the fully  
O-centered  and fully Ni-centered stripes are presented 
in Figs.3b and Fig.3c, respectively.  Although both the charge and spin 
along the corresponding O or Ni-centered stripes vary  
continuously, it is illustrative to define an "order parameter"  
which can clearly identify the 
regimes.  Here we define an order parameter as 
the average difference of spin density on the two nickel sites  
neighboring the  oxygen site belonging 
to the  O-centered stripe:  

\begin{equation}
\Omega\equiv <S_{i-1,j;left O-center}^{z}-S_{i+1,j;right O-center}^{z}> .
\end{equation}

In Fig. 3d we see that $\Omega$ exhibits a rapid crossover 
in the transition region, as compared to the variations in 
Fig. 3b, c, and is therefore useful for discriminating between 
O- and Ni-centered stripes. This is due to the positive 
feedback localizing charge and lattice distortion, and 
quenching spin.  
In the O-centered stripe region added holes are  
more accumulated along the oxygen stripe, while the spin density on 
the neighboring nickel sites are quenched, as some  
holes are also partially distributed on them; 
the spin density on those sites is 
smaller than that of other nickel sites far from the 
stripe, but the difference of the spin density on  
the two Ni sites adjacent to the O-centered stripe is almost 
zero. In a Ni-centered stripe, the holes accumulated on the Ni sites 
at the stripe center quench the spin density, but a large spin 
density is recovered on neighboring Ni sites. 

\vskip 6pt

\begin{figure}
\vskip -10pt
\centerline{
\epsfxsize=7.0cm \epsfbox{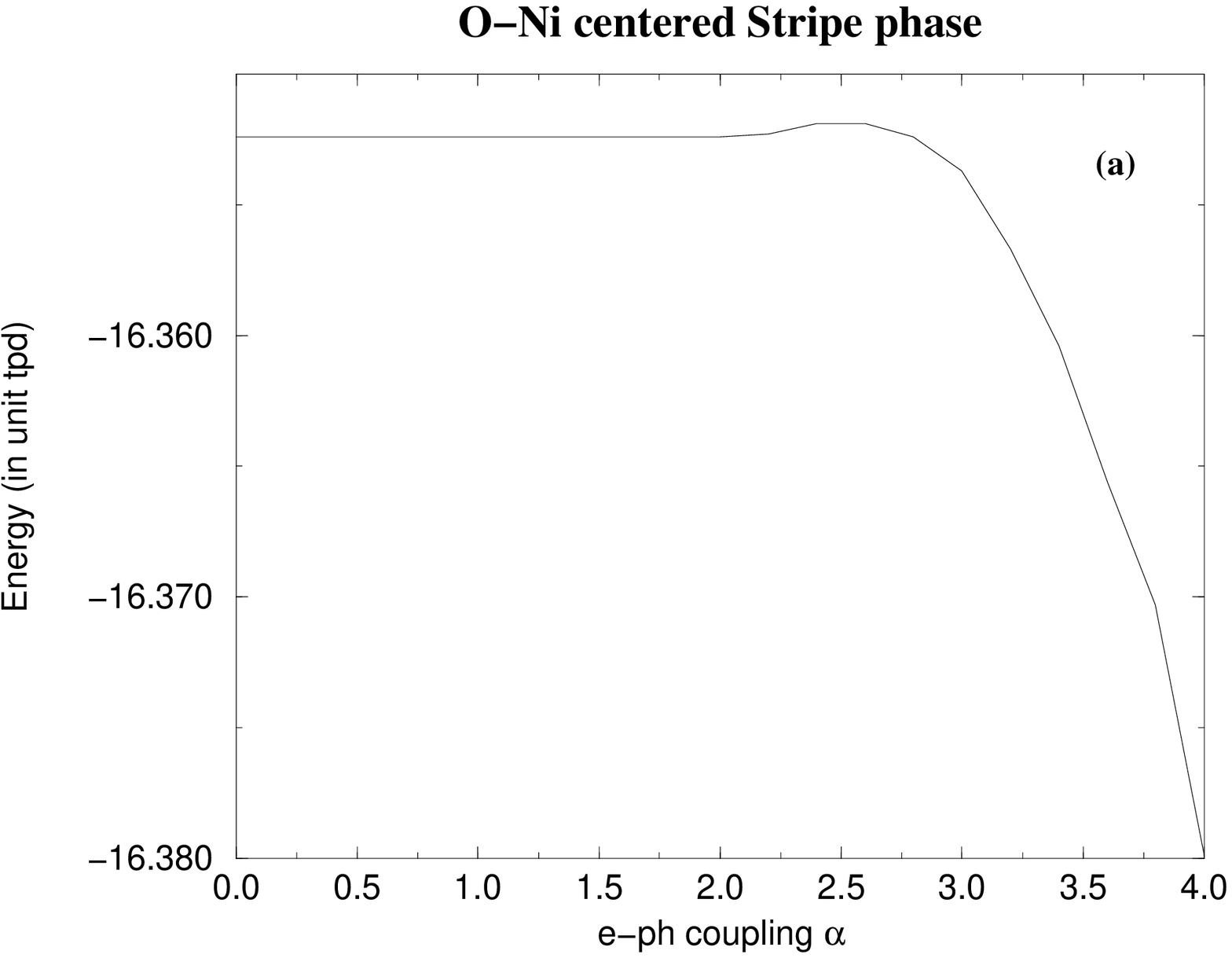}}
\centerline{
\epsfxsize=7.0cm \epsfbox{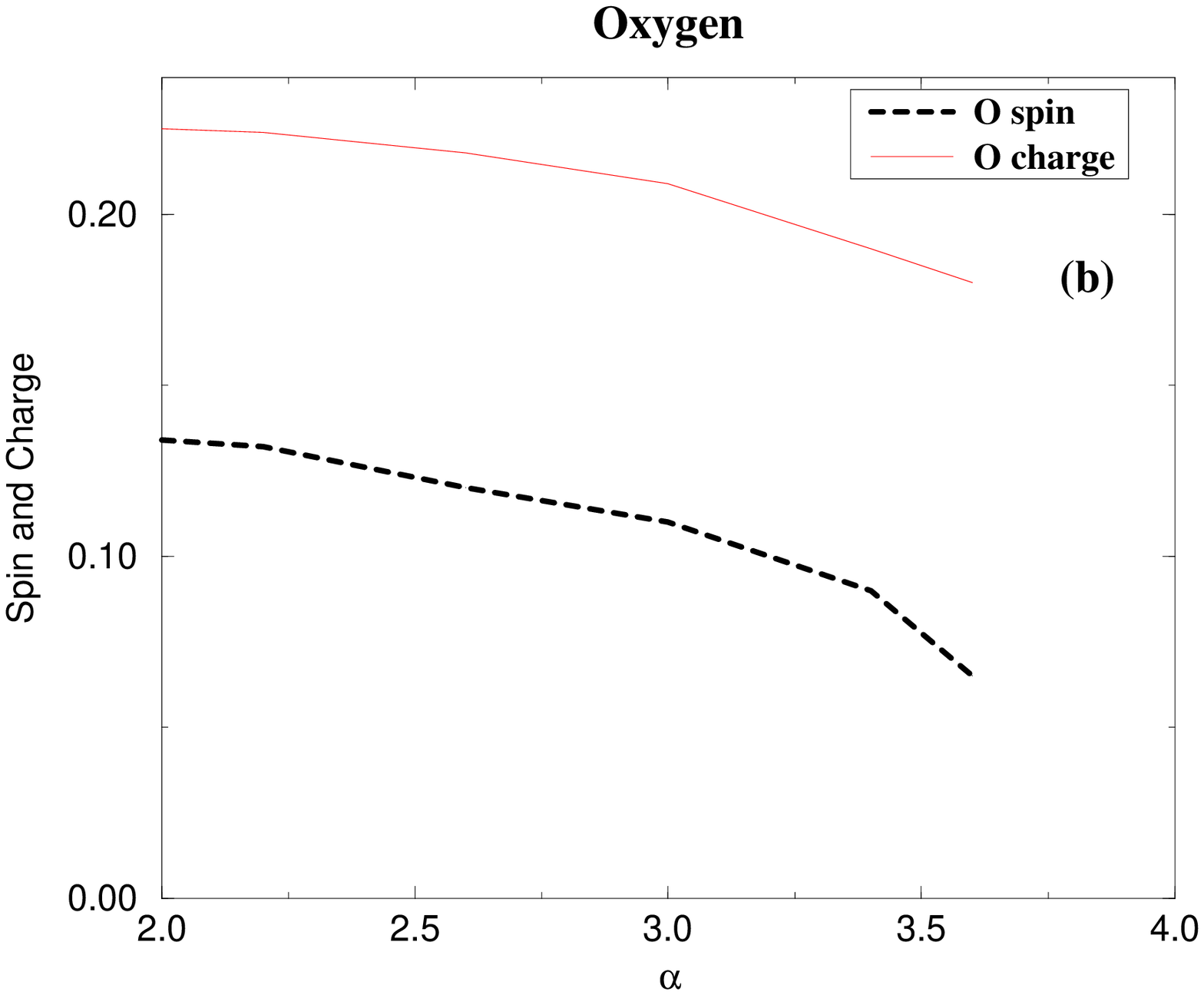}}
\centerline{
\epsfxsize=7.0cm \epsfbox{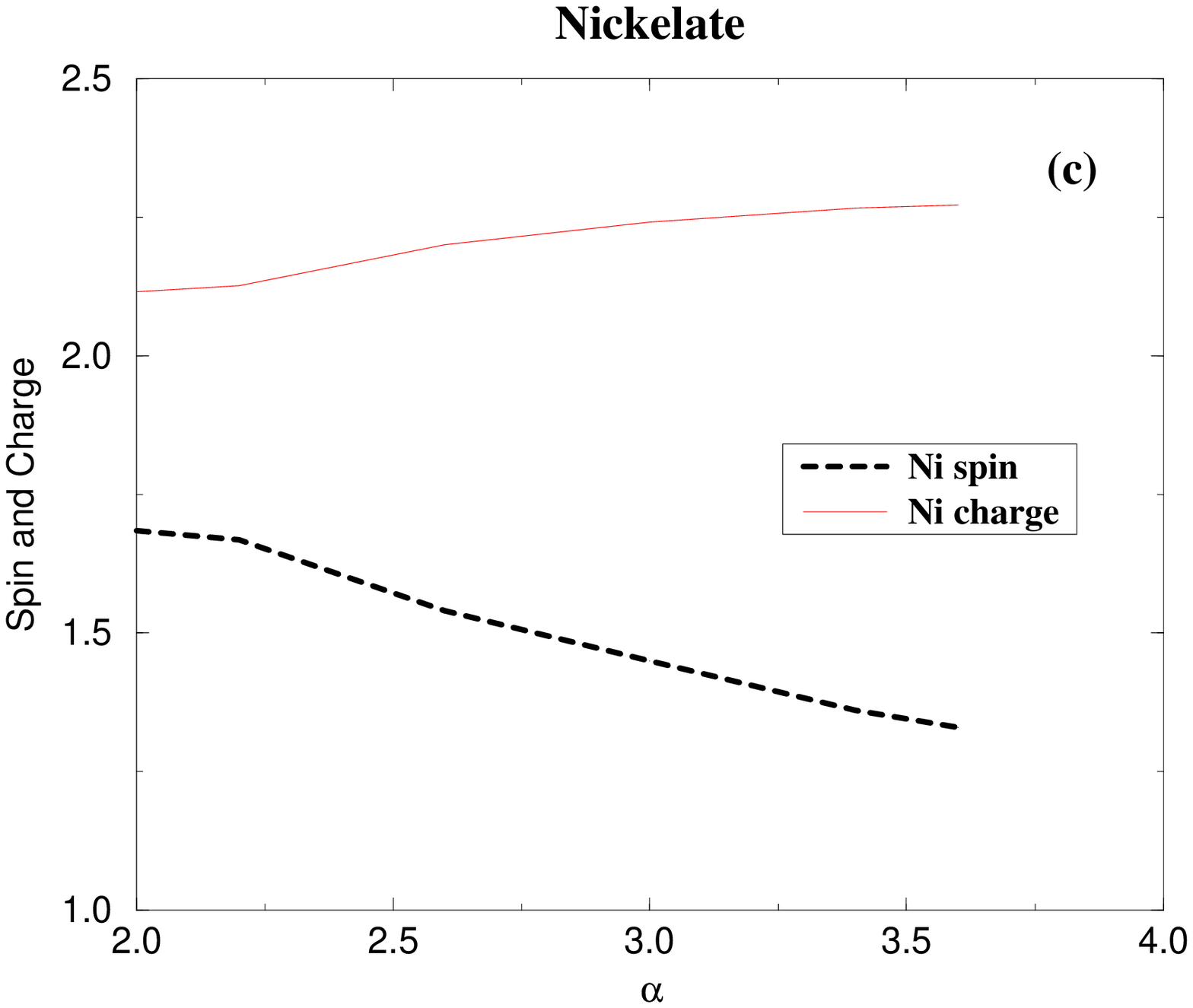}}
\centerline{
\epsfxsize=7.0cm \epsfbox{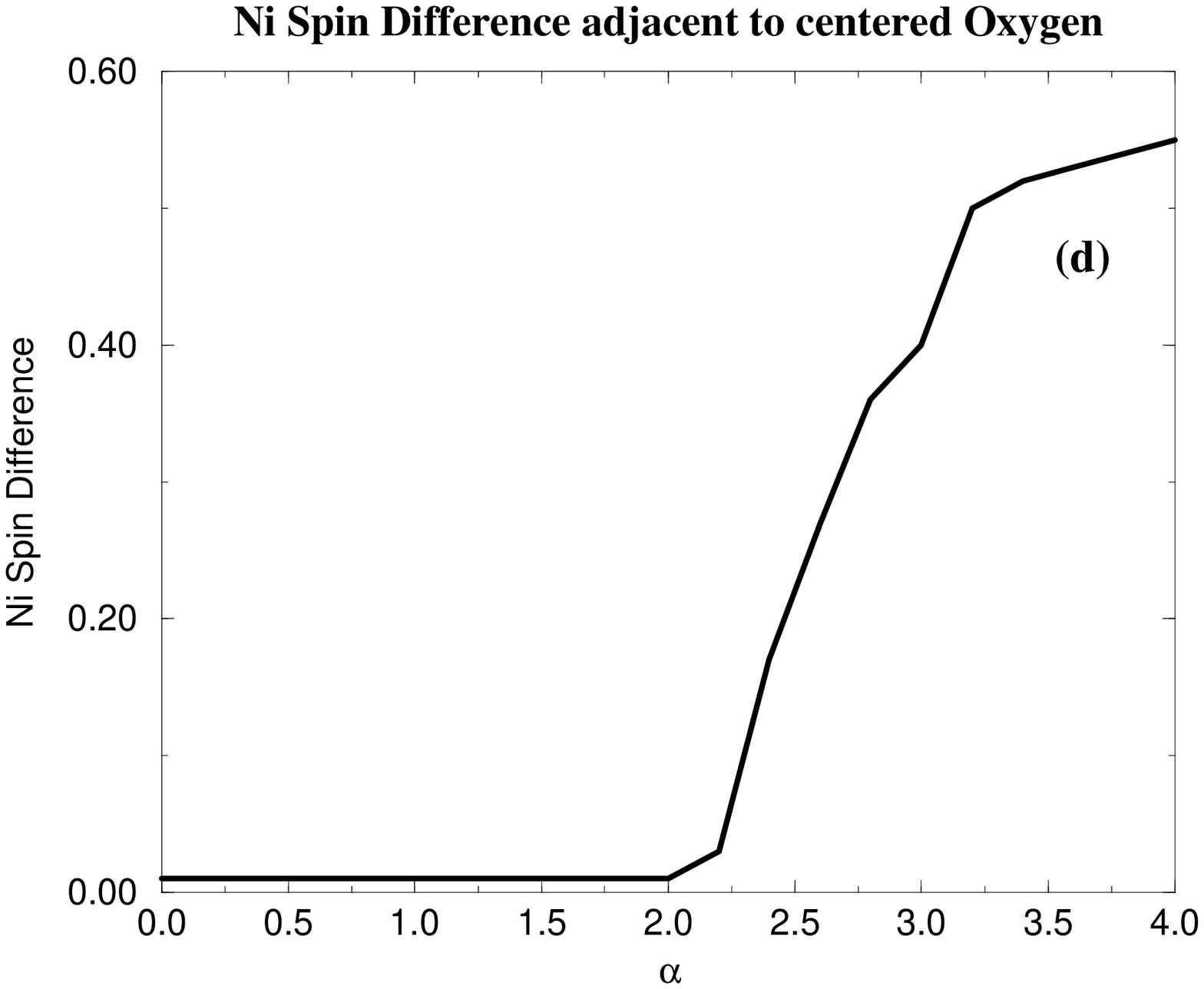}}
\vskip -10pt
\caption{Transition between the O-centered and  
Ni-centered stripe phases. (a) energy vs. e-l coupling , 
(b) and (c) variation of the spin and charge on  
oxygen stripes and nickel stripes, respectively, and (d) 
order parameter defined to  discriminate between the two 
stripe phases(see text).}
\end{figure}

Some recent experiments\cite{8} suggest that the Ni-centered stripe is more 
favored at low temperature, while the O-centered one is more favored 
at higher temperature. We have not included entropic effects  
in the present calculations which will contribute to the 
balance of stripe densities. However, as the e-l coupling constant is 
dependent on both the pressure and temperature, we suggest that 
a temperature dependent e-l coupling $\alpha$ could explain 
this behavior,   
because the elastic constant  should be stiffer at lower  
temperatures.  We can also expect that 
a pressure experiment would test the stripe phase 
dependence on $\alpha$, as the e-l interaction is likely to be even more 
sensitive to pressure than temperature. 

%\newpage

\section{signatures of the Ni-centered  
and  O-centered stripe phases}
Present neutron scattering experiment on the nickelate oxides 
do not unambiguously determine whether the 
stripe phase is Ni-centered or O-centered\cite{17}, 
so it is  necessary to predict more observable phenomena to 
discriminate between these two kinds of stripe. 
Experiments such as NMR, IR spectra, etc. are 
attractive candidates to  give information on the 
stripe type. Thus we give the predicted 
results on: (a) phonon spectra, in which 
different local modes(IR and Raman) are found to 
vibrate in the vicinity 
of the O-centered stripe and the Ni-centered one, respectively; 
(b) the IR spectra, which further identify the distinct localized 
features around the different stripe phases for the localized 
modes, and the corresponding optical absorption spectra 
on electronic energy scales from the local intra-gap levels; (c) magnetic 
excitations for both the Ni and  
O spin, since nuclear 
quadrupole resonances(NQR) and inelastic neutron  
experiments can give  
local information on magnetic excitations;  
and (d) charge excitation spectra associated with the respective 
stripes. 

\subsection{Local phonon modes and the IR spectra}
After forming either Ni-centered or O-centered stripe 
phases, the inhomogeneity of the systems suggests that {\it local}  
phonon modes may be self-consistently induced as a result of the 
locally renormalized spring-constants 
\cite{19}, which is calculated using the renormalized 
spring-constant matrix(for e.g., see Appendix H in Ref.\cite{19}).  
This is indeed the case. 
The phonon dispersions calculated for the 
stripe phases are illustrated in Fig.4a for O-centered and 
Fig.5a for Ni-centered stripe phases. 
In both cases, the phonon 
modes are strongly softened due to the lattice distortions, around 
the stripes induced by the e-l interaction. 
When the e-l interaction is present, 
local phonon bands form for both O-centered and Ni-centered 
stripes. It is clearly seen that more localized and softened  
phonons are found in the Ni-centered stripe phase, as this 
needs stronger e-l interaction.  

\begin{figure}
\vskip -10pt
\centerline{
\epsfxsize=7.0cm \epsfbox{empty.eps}}
\centerline{
\epsfxsize=7.0cm \epsfbox{empty.eps}}
\vskip -10pt
\caption{Phonon dispersion (a) and lowest frequency localized 
phonon mode vibration pattern in an O-centered stripe phase case (b). 
Arrows represent direction of the vibrating oxygens, 
and the length is proportional to the vibrational amplitude.}
\end{figure}

\begin{figure}
\vskip -10pt
\centerline{
\epsfxsize=7.0cm \epsfbox{empty.eps}}
\centerline{
\epsfxsize=7.0cm \epsfbox{empty.eps}}
\vskip -10pt
\caption{Phonon dispersion (a) and lowest frequency localized 
phonon mode vibration pattern in a Ni-centered stripe phase case (b). 
Arrows represent direction of the vibrating oxygens, 
and the length is proportional to the vibrational amplitude.}
\end{figure}

With our parameters,  the lowest 
phonon modes for the O-centered stripe phase with $\alpha=2.0$ for 
a 5x5 system is 481.7cm$^{-1}$, and the corresponding lowest 
phonon mode  
for the Ni-centered stripe phase when $\alpha=3.6$ is 426.5cm$^{-1}$. 
These two lowest modes and the associated eigenvectors  
for the two respective stripe 
phase are illustrated in Figs.4b and 5b. 
The arrows represent the oxygen displacement patterns, 
and the length of the arrows is proportional 
to the amplitude of the vibrational modes. We  
see that these lowest phonon modes are 
localized around the  O-centered stripes or 
the Ni-centered stripes, respectively.  The localized phonon 
modes and their corresponding distortions in the vicinity 
of stripes suggest  observable signatures 
in an IR absorption experiment.

\begin{figure}
\vskip -10pt
\centerline{
\epsfxsize=7.0cm \epsfbox{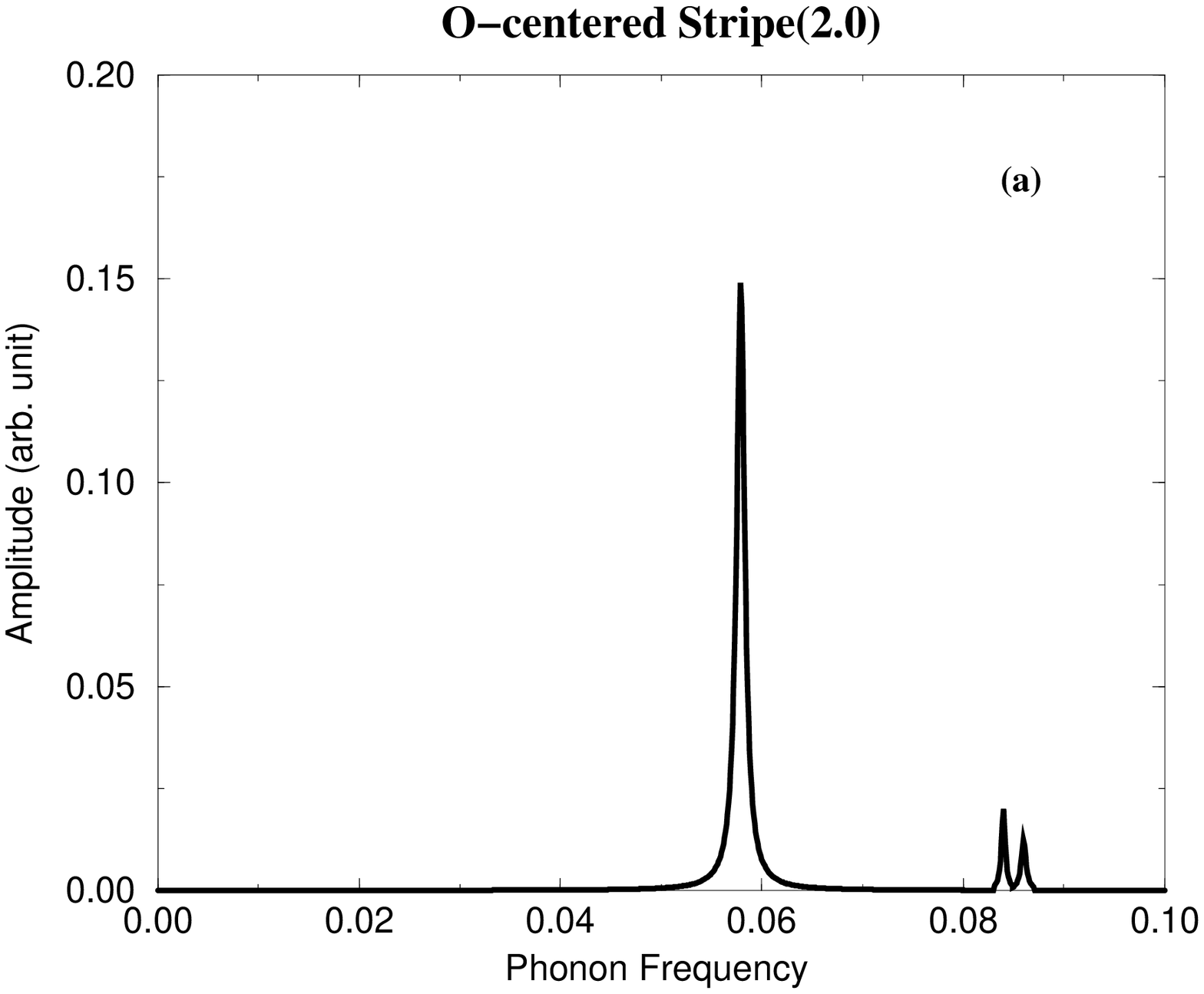}}
\centerline{
\epsfxsize=7.0cm \epsfbox{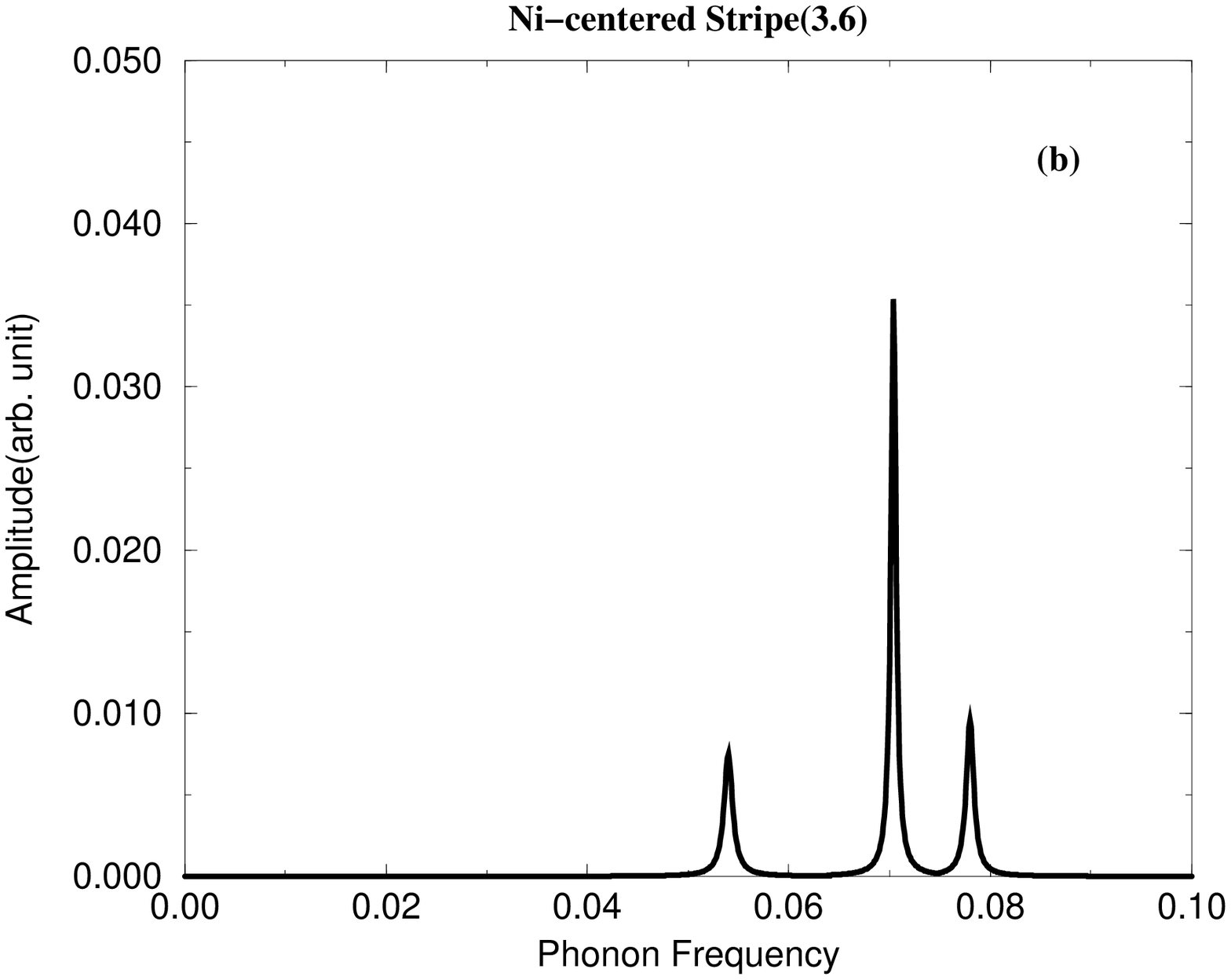}}
\vskip -10pt
\caption{IR spectra of a 3x3 system for O-centered 
stripe phase (a) and Ni-centered stripe phase (b),  
respectively.}
\end{figure}

\begin{figure}
\vskip -10pt
\centerline{
\epsfxsize=7.0cm \epsfbox{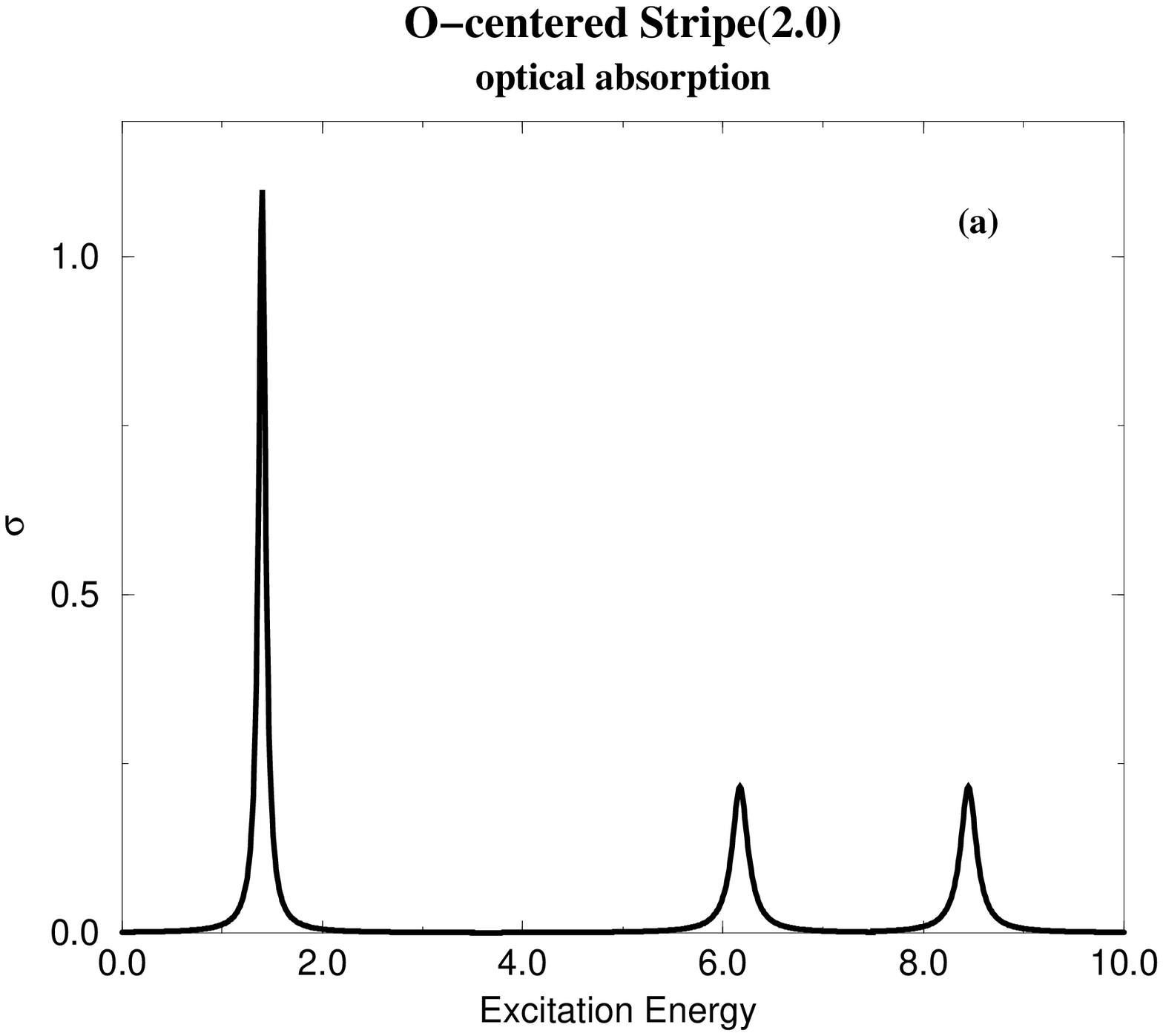}}
\centerline{
\epsfxsize=7.0cm \epsfbox{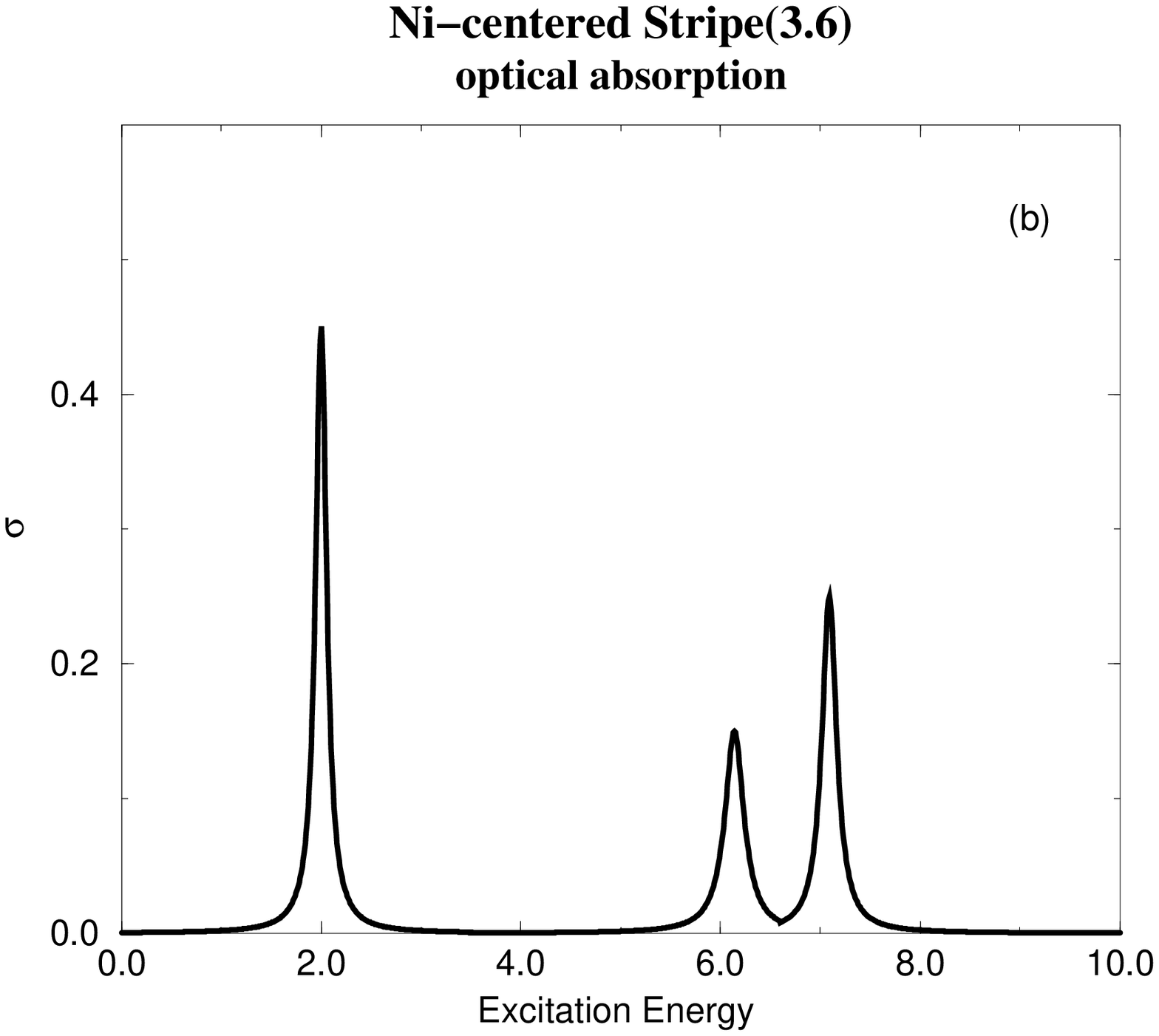}}
\centerline{
\epsfxsize=7.0cm \epsfbox{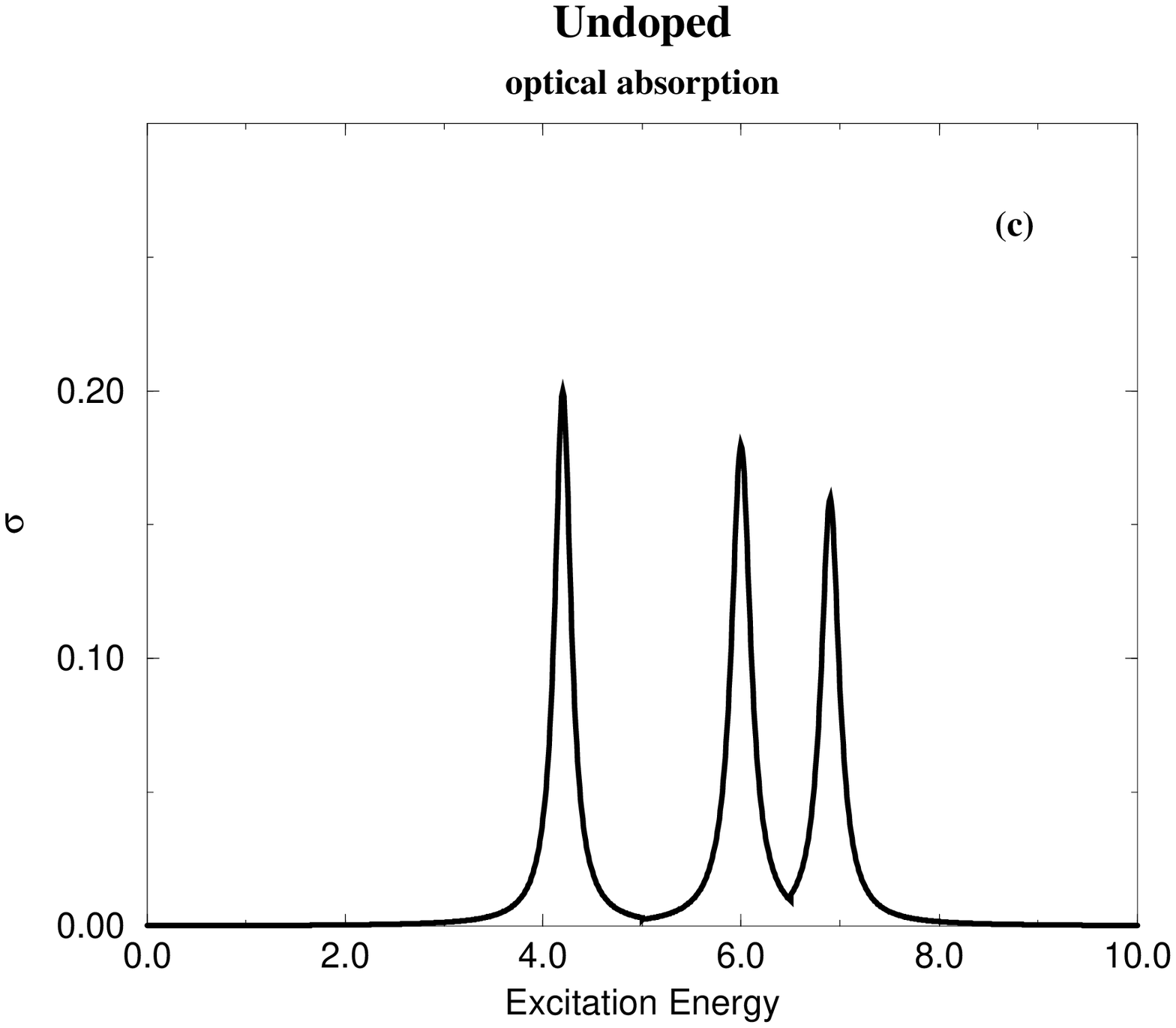}}
\vskip -10pt
\caption{Optical absorption of a 3x3 system on electronic 
energy scales for O-centered stripe phase (a), Ni-centered 
stripe phase (b), and undoped case (c), respectively.}
\end{figure}

In Fig.6a and Fig.6b, the IR spectra are plotted for an O-centered 
and a Ni-centered stripe, respectively, for a small 
3x3 system.(The RPA calculation needs large memory, 
so we use a smaller system). 
The three infrared peaks in Fig.6a arise from
the one  distinct localized mode at $\omega=0.0579$, and 
other less localized modes with frequencies $\omega=0.0840$ and 
$\omega=0.0860$. The lowest phonon mode corresponding to 
the first peak  and its lattice vibration pattern  
are clear  in Fig.4b, which illustrates the 
localized character around the corresponding O-centered 
stripe. The other two peaks corresponding to the two 
less localized phonon modes also confine the lattice vibrations  
mainly around the stripe phase. The three 
peaks in Fig.6b for the Ni-centered stripe phase come  
from three localized phonon modes at $\omega=0.054, 0.0704$, and 
$0.0780$, respectively. The stronger electron-lattice interaction 
in the Ni-centered stripe phase results in  
the more strongly localized nature of these 
phonon modes. Fig.5b illustrates the character of the  
localized phonon modes in this case. We emphasize that, since the 
Ni-centered stripe phase is favored in the strong e-l  
coupling region, they will be accompanied by lower frequency and 
more localized phonon 
modes. 

As there are two orbitals($d_{x^2-y^2}$, $d_{3z^2-1}$) on each nickel site, 
local defects  
induced by orbital mixing can have effects in our 
finite system. We find that the hole occupation of the two 
orbitals is quite different along the stripe, especially 
in two corners in Fig.1b and Fig.5b. To emphasize the finite 
size effect by the orbital mixing, we explored the larger 
7x7 system and find this effect is much reduced, especially around the 
two corners along the stripe. This orbital mixing 
effect is not our focus in this 
paper, and the main physics we are interested here is not sensitive 
to it. Details on orbital mixing will be reported elsewhere. 

The optical absorption spectra for both of these 
two stripe phases and the undoped case are illustrated 
in Figs. 7a, 7b and 7c on 
the electronic energy scale. The peaks correspond to many 
electron-hole excitations from the filled bands to unfilled 
bands, respectively.   
There are additional localized phonon modes around the stripes which are 
raman active. It would be important to excite these in resonant 
raman studies by tuning into the electronic gap states(Fig.2 ) associated 
with the corresponding stripes. Such excitation profiles would 
provide very strong correlations between high- and low-energy characteristics 
of specific stripes\cite{21,22}.

\subsection{Magnetic and Charge Excitation spectra}
Since NMR and NQR experiments can accurately probe  
local magnetic excitation, we have also calculated the 
RPA magnetic(as well as charge) excitations 
for both the Ni and the O sites. As will be 
shown below, the spectral weight of the magnetic excitations  
again contains specific features characteristic of the stripes. 

The spectral weight for particle-hole and phonon excitations 
is given by the imaginary part of the two body Green's 
function in each channel, $A$:  

\begin{equation}
Im\chi_{A}({\bf q}, \omega)=\displaystyle\frac{\pi}{N_{cell}}\sum
\limits_{n\neq 0}|<0|\hat{A}({\bf q})|n>|^{2}
\delta(\omega-(E_{n}-E_{0})). 
\end{equation}
Here  $A$ is the operator whose spectral weight we wish  
to obtain. For magnetic excitations, we assign $A$ as the 
transverse components of the spin 
\begin{equation}
{\bf S_{Ni~ or ~O}}({\bf q})=
\sum\limits_{{\bf r}\in Ni~ or ~O}e^{i{\bf q}\cdot {\bf r}}
\frac{1}{2}{\bf \sigma_{r}}
\end{equation} 
with ${\bf \sigma_{r}}=\sum\limits
_{\tau, \tau^{'}}c_{r,\tau}^{\dagger}{\bf \sigma_{\tau, \tau^{'}}}
c_{r,\tau^{'}}$, and ${\bf \sigma}$ standing for the Pauli spin 
matrix. For the amplitude of charge, we can assign a corresponding  
operator\cite{19}. Here $N_{cell}$ is the number 
of NiO$_{2}$ units. 

\begin{figure}
\vskip -10pt
\centerline{
\epsfxsize=7.0cm \epsfbox{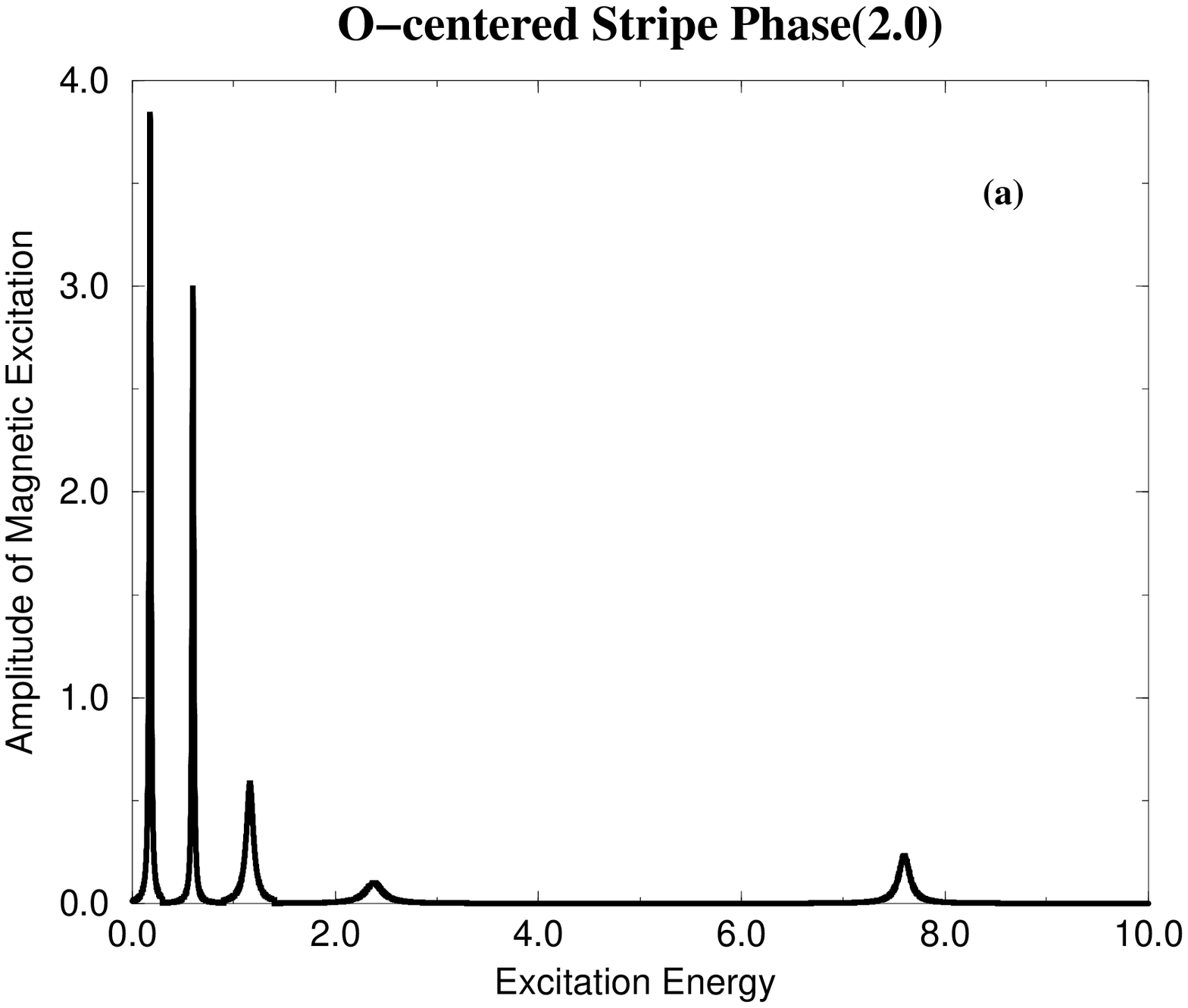}}
\centerline{
\epsfxsize=7.0cm \epsfbox{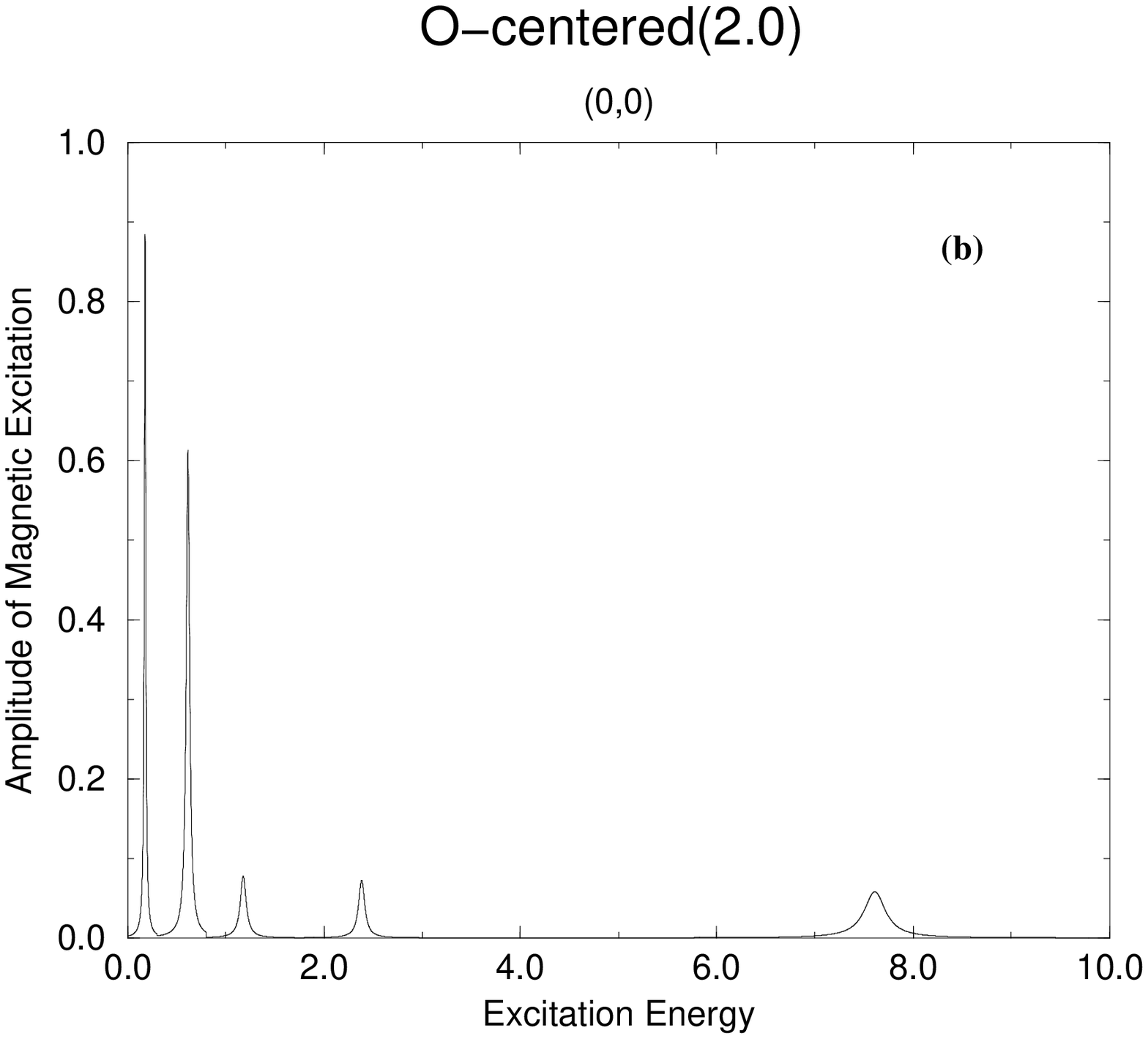}}
\centerline{
\epsfxsize=7.0cm \epsfbox{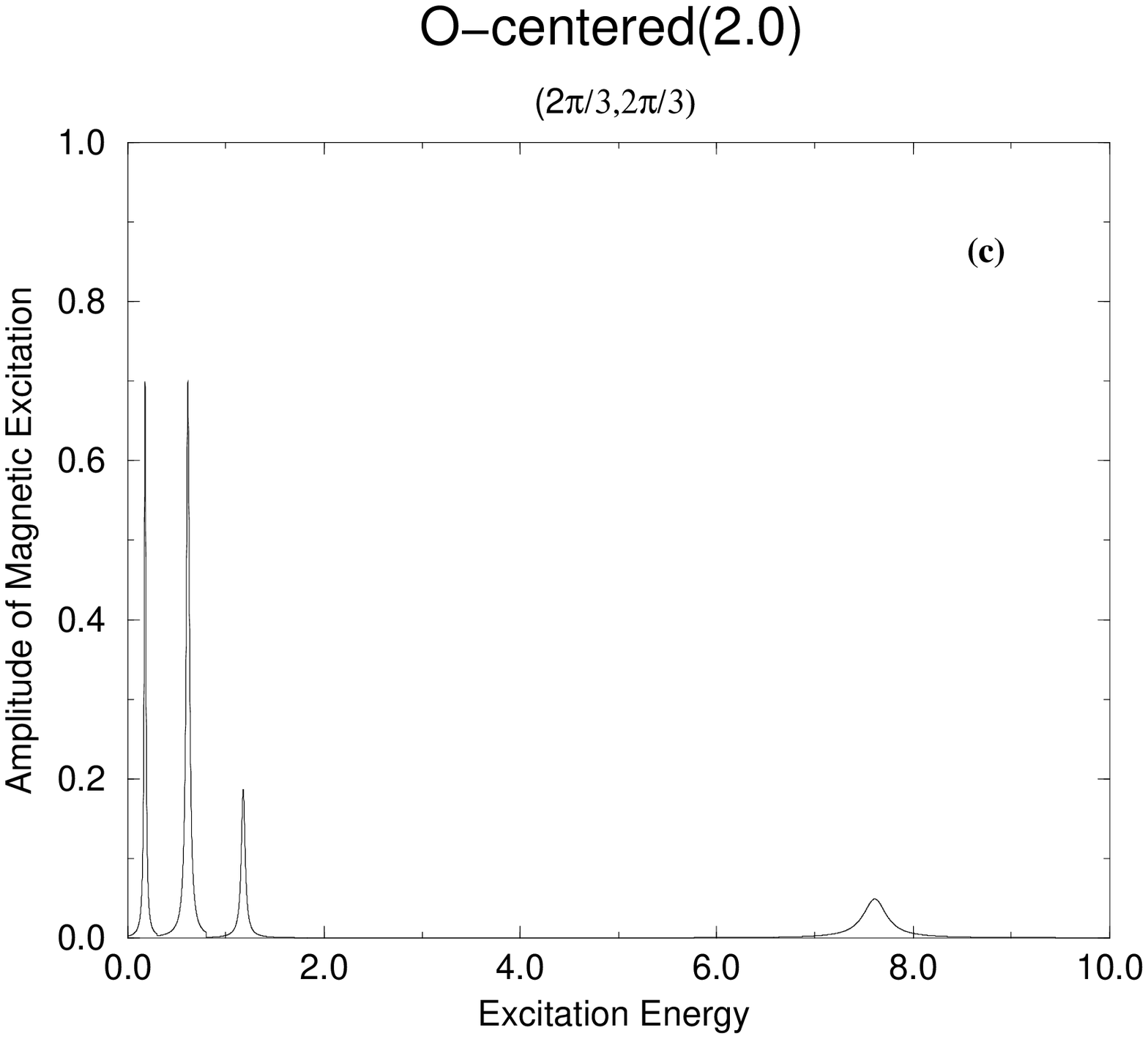}}
\vskip -10pt
\caption{Magnetic excitation spectra of nickel  
sites for an O-centered stripe phase. (a) is the sum over  
all the momenta. (b) and (c) are the magnetic excitations  
at specific momenta, as identified in the figure.}
\end{figure}

The magnetic excitations for all the nickel sites in an O-centered 
stripe phase are illustrated in Figs.8a, 8b, and 8c.  
Fig.8a represents the sum over all the momenta ${\bf q}$, 
and  can be  
compared to  NMR experiments. Fig.8b and Fig.8c 
correspond to the spectral weight of magnetic excitation on 
Ni sites at specific momenta, here ${\bf q}=(0, 0)$ for 
Fig.8b and ${\bf q}=(2\pi/3, 2\pi/3)$ for Fig.8c. This 
information is accessible to magnetic inelastic neutron scattering.  
Five distinct 
peaks appear at $\omega=0.178, 0.60, 1.16, 2.38$ and 
$7.6$, of which  
$\omega=0.178, 0.60$ 
have large amplitudes. Detailed 
analysis of the origins of these two peaks reveals very specific localized 
stripe information. For the $\omega=0.178$ peak,  the 
RPA wave function corresponds to  electron-hole 
excitation of the localized state in the charge transfer gap of 
the corresponding undoped system. 
By checking the wave function of the two localized 
electronic states, labelled as L1 and L2 in Fig.2a,  
we find  that  both electronic 
states are localized around the stripe: i.e., the 
distribution of the wavefunction amplitude is mainly around the stripe 
sites. Excitation between the two localized states  
 represents local spin-flips of some oxygens on the O-centered 
stripe. Thus the new low-lying magnetic excitations clearly 
reflect  information localized 
around the stripe. A similar conclusion is  obtained 
for the second largest peak $\omega=0.60$.  

\begin{figure}
\vskip -10pt
\centerline{
\epsfxsize=7.0cm \epsfbox{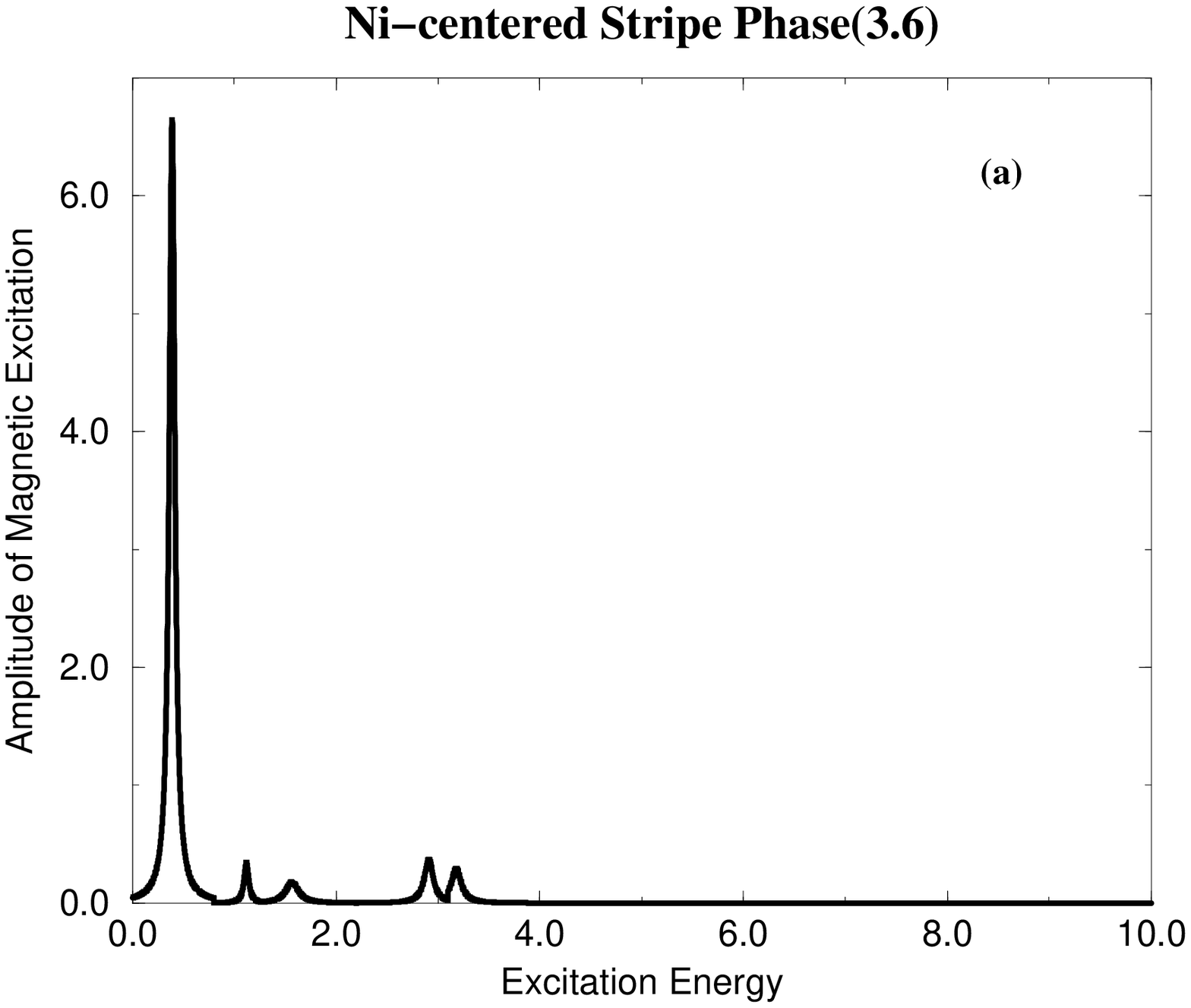}}
\centerline{
\epsfxsize=7.0cm \epsfbox{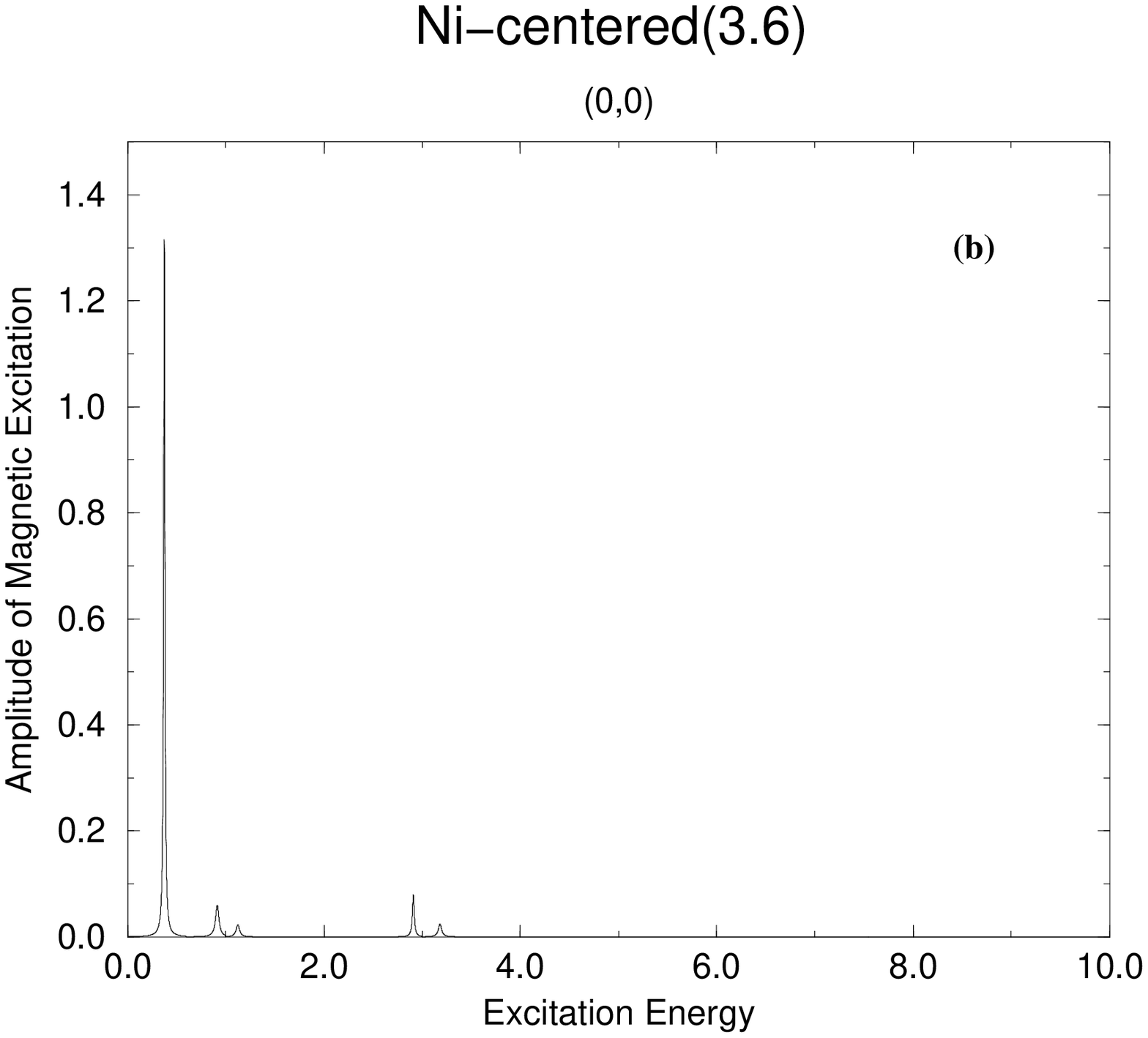}}
\centerline{
\epsfxsize=7.0cm \epsfbox{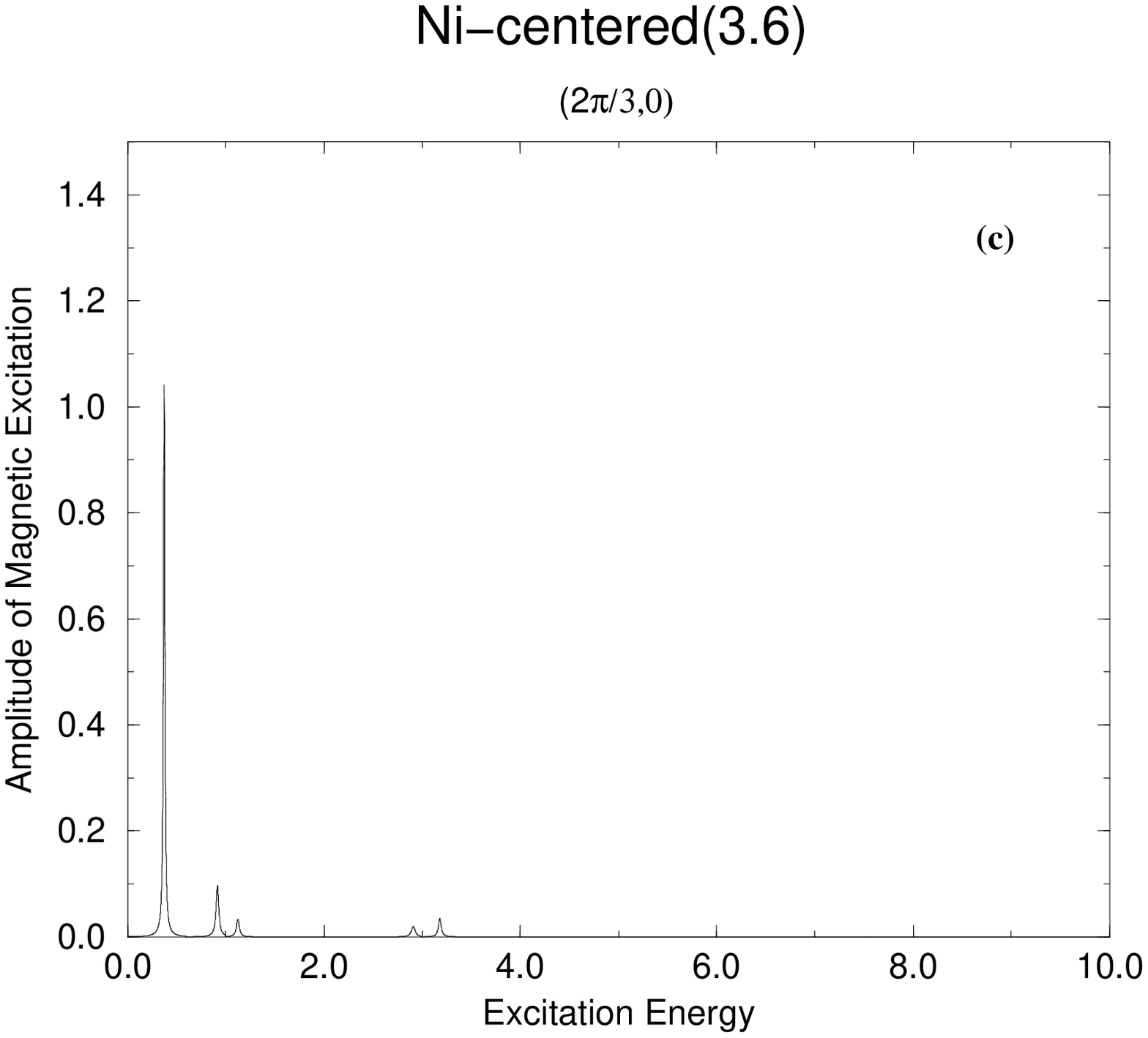}}
\vskip -10pt
\caption{Magnetic excitation spectra of nickel 
sites for a Ni-centered stripe phase: (a) the sum over  
all the momenta. (b) and (c) the magnetic excitations 
at specific momenta, as identified in the figure.}
\end{figure}

\begin{figure}
\vskip -10pt
\centerline{
\epsfxsize=7.0cm \epsfbox{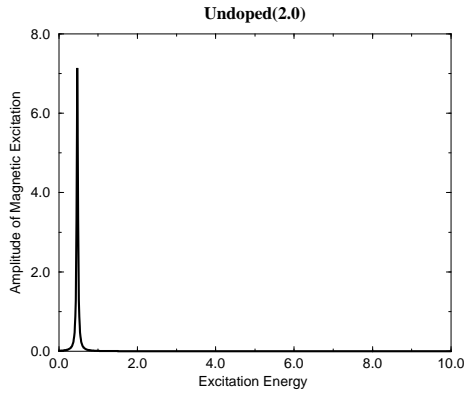}}
\vskip -10pt
\caption{Magnetic excitation spectra of nickel  
sites for the undoped case, summed over all the momenta.}
\end{figure}

For the magnetic excitations of the nickel sites in the 
Ni-centered stripe phase, the spectral weight is represented 
in Figs.9a, 9b, and 9c, respectively. Similar to the O-centered 
case, Fig.9a is the sum over all the momenta and can be compared  
to an NMR experiment, while Figs.9b and 9c correspond to  
momenta at ${\bf q}=(0,0)$ and ${\bf q}=(2\pi/3,0)$. Five peaks 
are found: at $\omega=0.39, 1.12, 1.568, 2.91$ and $3.18$. 
Here we see a distinct difference between the Ni-centered 
stripe phase and the O-centered one: only one main peak with 
large amplitude appears, at $\omega=0.39$. This 
peak again reflects  information localized around the Ni-centered 
stripe. By checking the wavefunction of the main  excitation between 
the two localized electronic states,  which are mainly localized  
around the Ni-centered stripe, the magnetic excitation 
in this case is found to come 
from  spin-flips of the oxygens neighboring the Ni-centered 
stripe. This is different from the O-centered case above. 
Thus both of the magnetic excitation spectra are characteristic of 
their corresponding stripes,  and give distinct features, which 
we expect to be observable in NMR, NQR or inelastic 
neutron scattering experiments.   
For reference, we present results for the undoped case in Fig.10. Here,  
the peak at $\omega=0.48$ corresponds to the electron-hole 
excitation across the charge-transfer-gap. It is clearly seen that 
the intragap electronic levels for the stripe phases induce additional 
peaks, as illustrated in Figs. 8 and 9. 

\begin{figure}
\vskip -10pt
\centerline{
\epsfxsize=7.0cm \epsfbox{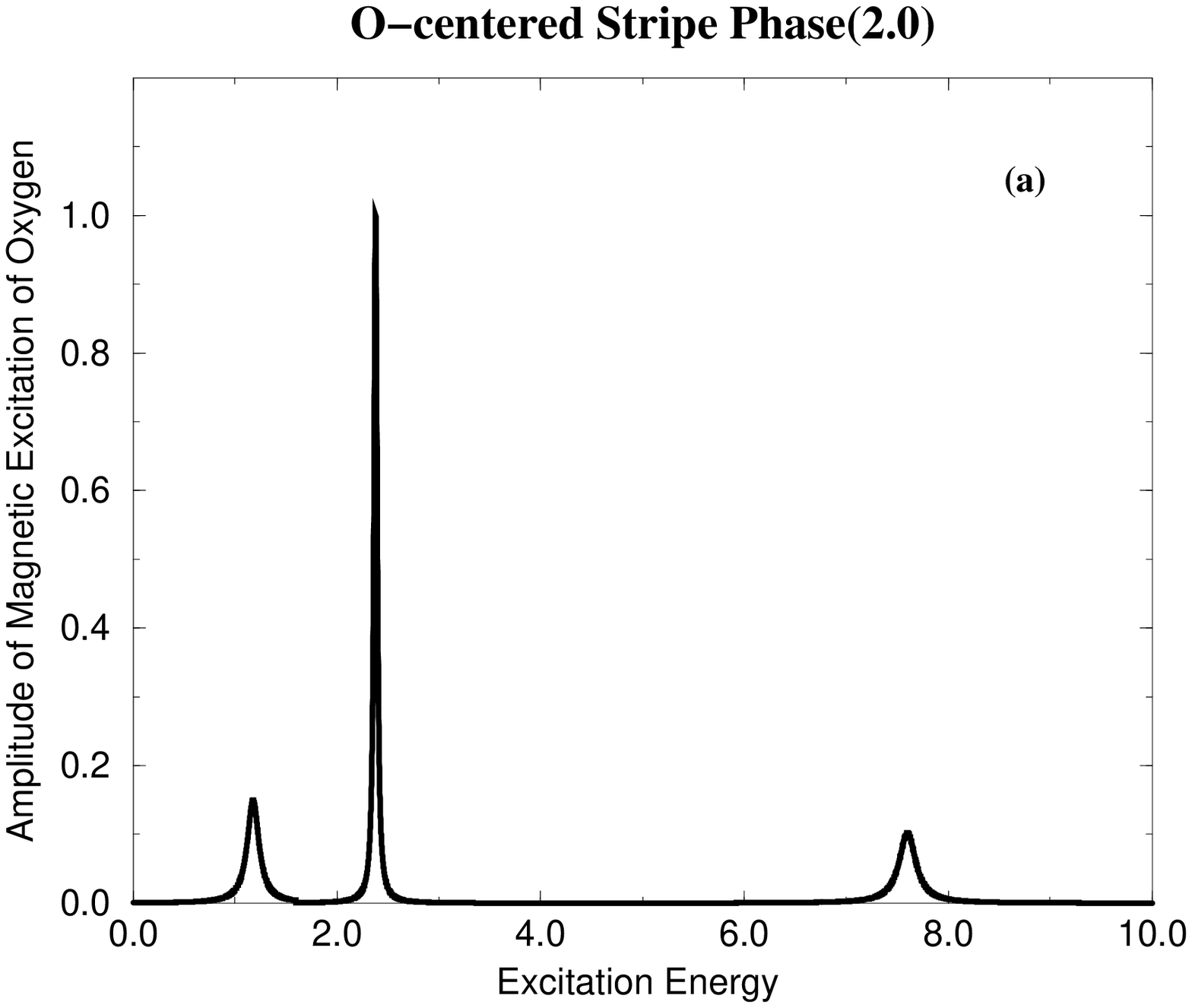}}
\centerline{
\epsfxsize=7.0cm \epsfbox{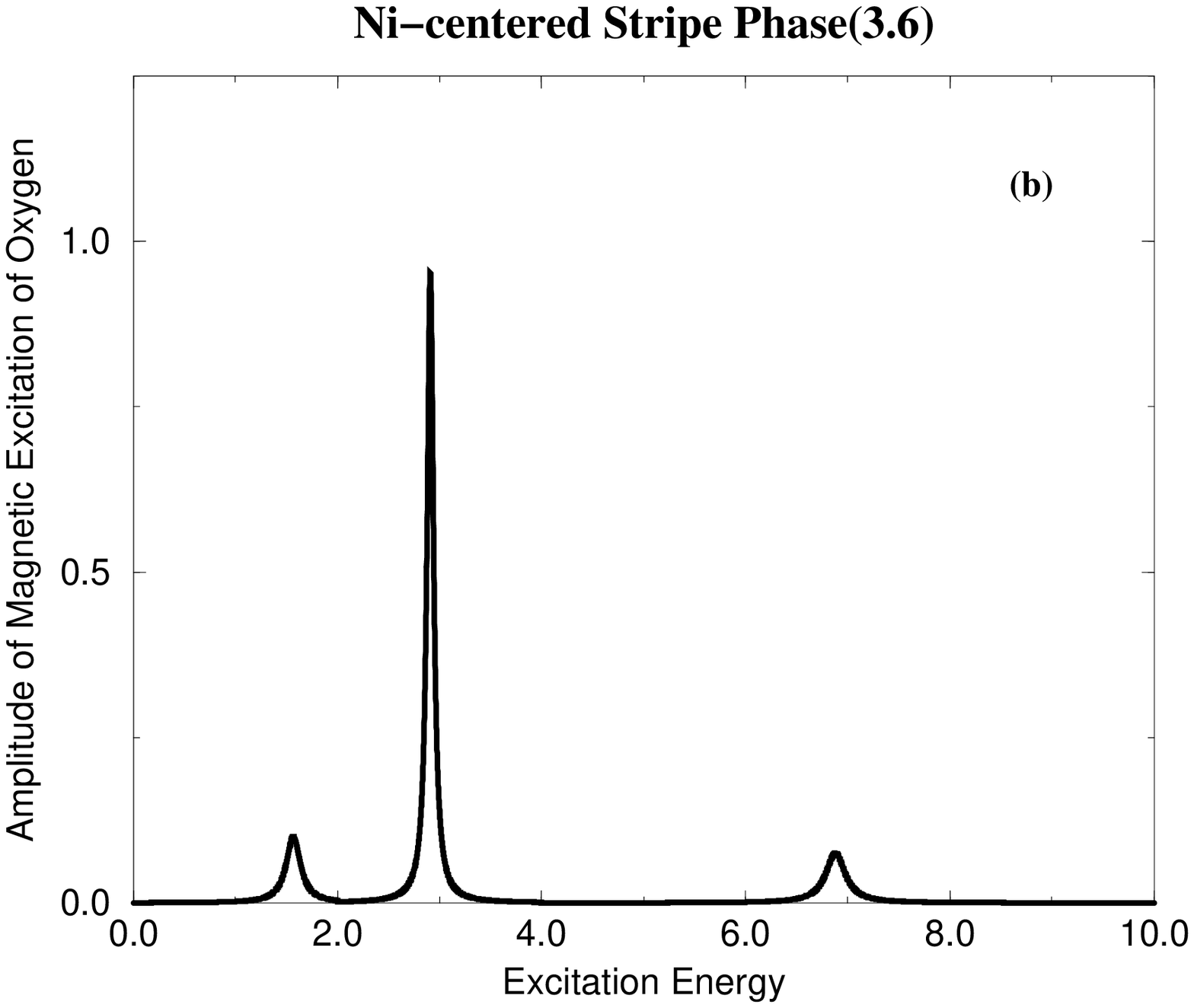}}
\centerline{
\epsfxsize=7.0cm \epsfbox{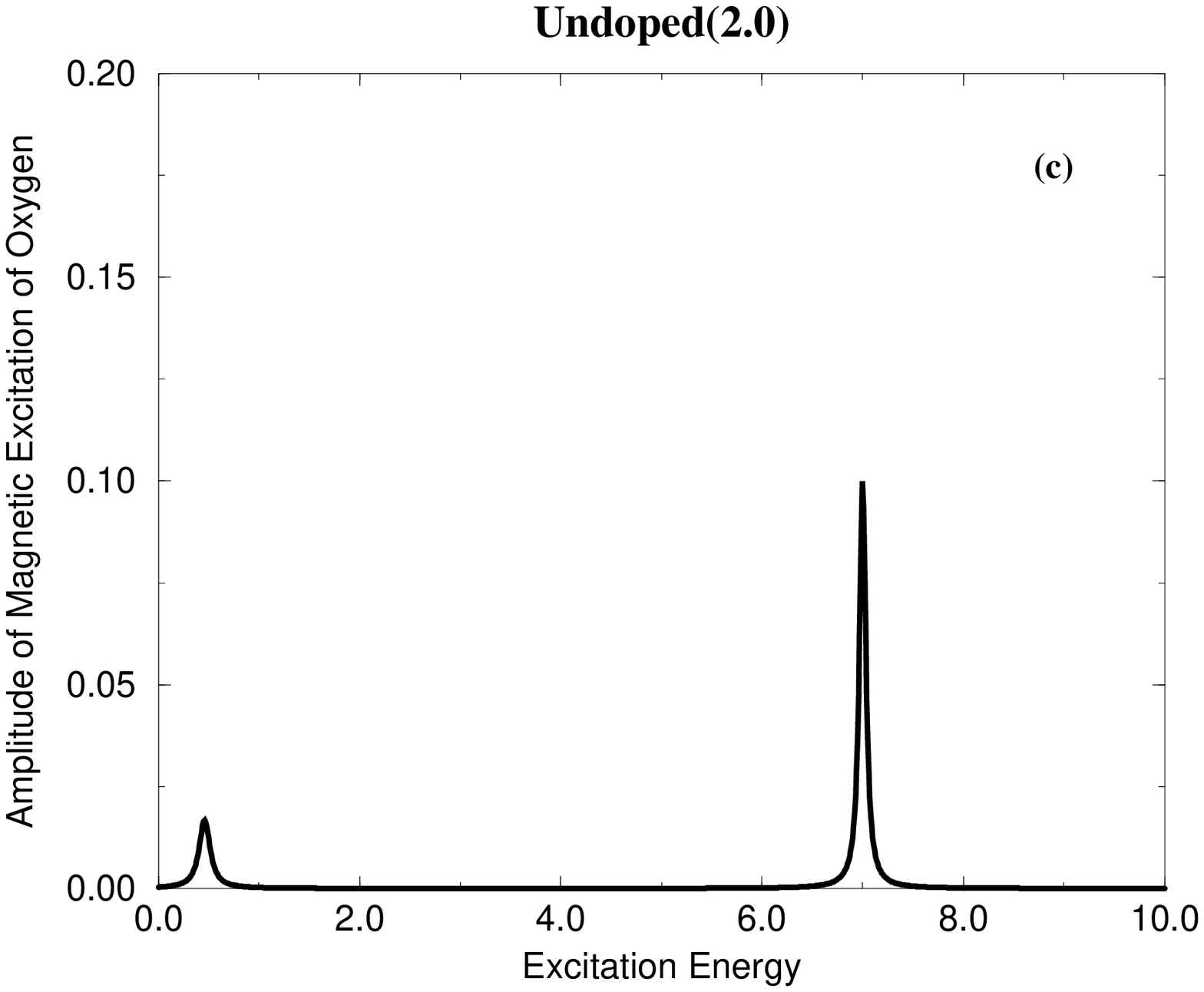}}
\vskip -10pt
\caption{Magnetic excitation spectra of oxygen 
sites for an O-centered stripe (a), a Ni-centered 
stripe (b), and the undoped case (c), summed over 
all momenta.}
\end{figure}

\begin{figure}
\vskip -10pt
\centerline{
\epsfxsize=7.0cm \epsfbox{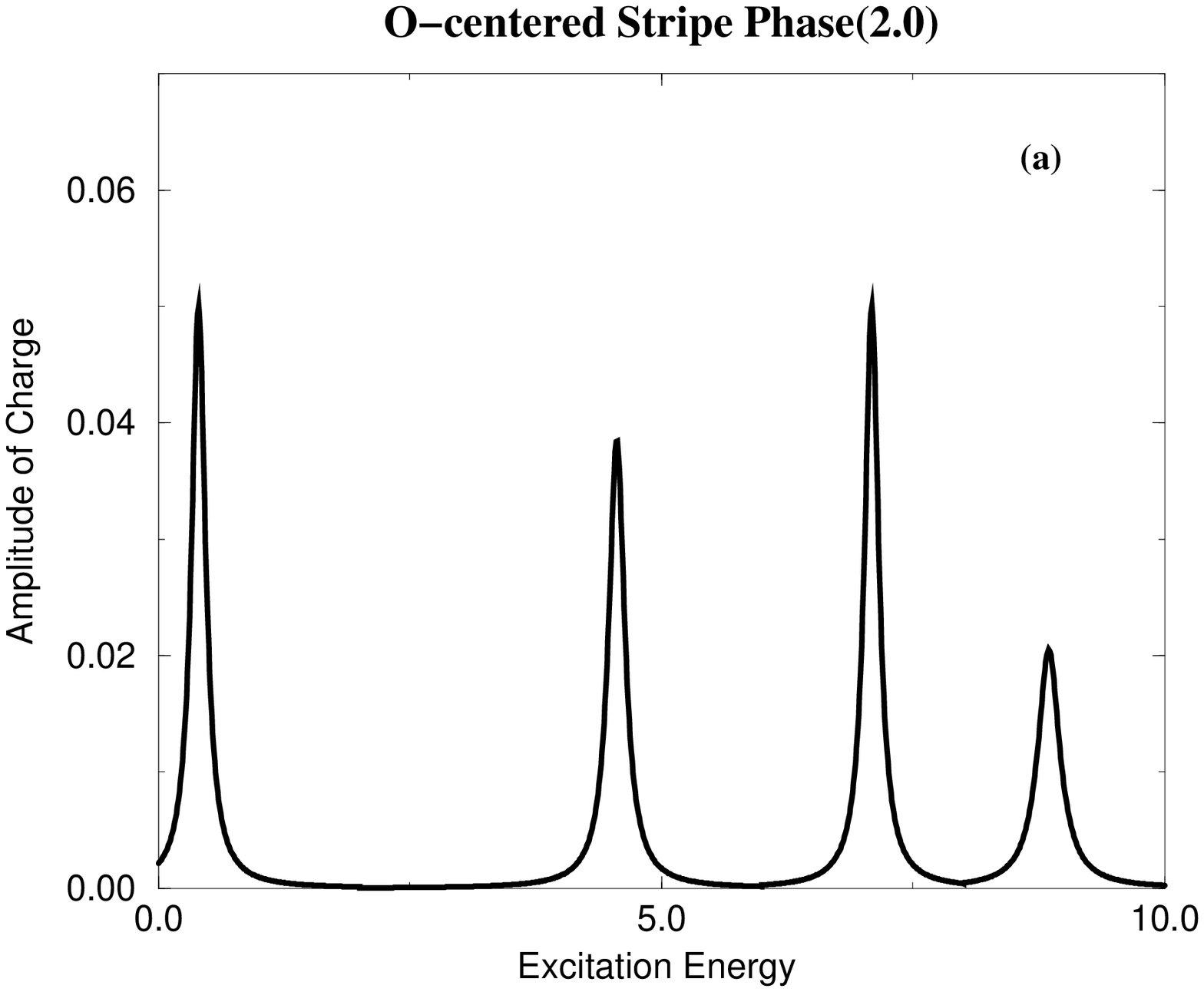}}
\centerline{
\epsfxsize=7.0cm \epsfbox{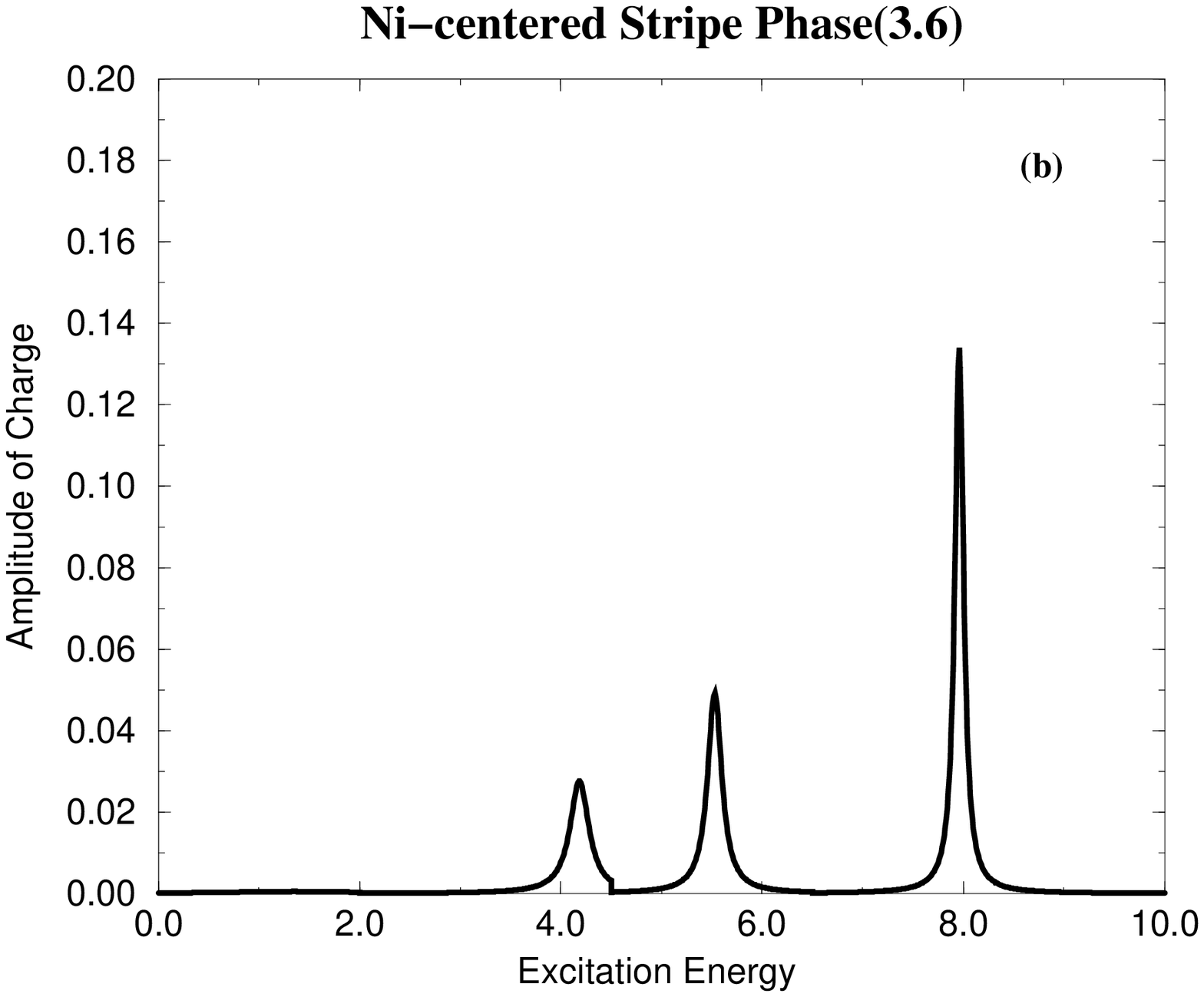}}
\vskip -10pt
\caption{Amplitude of charge excitations  of nickel   
sites for an O-centered stripe (a), and a Ni-centered 
stripe (b).}
\end{figure}

In Figs.11a, 11b, 11c, we also illustrate  results for 
the magnetic excitation 
spectra of the {\it oxygen} sites for an 
O-centered stripe phase, Ni-centered stripe phase and 
undoped case, respectively. Due to the small spin on 
oxygen sites, the amplitude is smaller 
than the corresponding signatures for the magnetic excitations on 
nickel sites.  The amplitude for {\it charge} excitations  on nickel   
sites are shown in Figs.12a and Fig.12b for the O-centered 
and Ni-centered stripe phases, respectively. As anticipated, 
the strength of these signatures is weaker than the magnetic 
excitations on the oxygen for the two kinds of 
stripe phases. This demonstrates the more localized 
character of the added holes on an O-centered 
or Ni-centered stripe, as the main feature  
of a stripe is  charge localization along the stripe,  
dividing the whole system into hole-rich and hole-poor 
regions.

%\newpage

\section{Conclusion}
In summary, we have studied nickel-centered  
and the oxygen-centered stripe phases, recently suggested in 
doped nickelate oxides, including both e-e and 
e-l interactions in a model of a single NiO$_{2}$ plane. 
A transition from an O-centered stripe 
phase to a Ni-centered one is found, as a function of the   
e-l interaction strength.  Due 
to the difficulty of neutron experiments unambiguously 
distinguishing specific 
stripe types,  we have provided  predictions of additional observable 
signatures which may be accessible to  
experiments: localized phonon modes, IR spectra, and 
the spectra for various charge and spin 
excitations, as well as the electronic optical absorption. 
These show novel and 
distinct features in the vicinity of the respective stripes. 
NMR or NQR experiments, together with inelastic neutron scattering, 
are very promising for verifying our 
predicted signatures and can give  clear local information.  
Likewise, resonant raman excitation profiles would be 
extremely valuable for confirming relationships between low- and 
high-energy signatures of stripes\cite{22}.  
Charge localization into mesoscale stripe patterns is of central 
interest in strongly coupled spin-charge-lattice materials.  
Thus, the discrimination between stripe types will be of great value  
for understanding charge localization and ordering mechanisms 
in nickelates and related complex electronic 
materials, such as cuprate and manganese oxides. 
We expect that the observable signatures 
we have predicted 
can help to focus a coordinated set of experiments to identify   
specific stripe phases. 

\section*{Acknowledgments}
We have benefitted from valuable discussions with P. C. Hammel,   
Y. Yoshinari and J. Zaanen. This work was supported 
by the U. S. Department of Energy.

\end{document}